\documentclass[11pt, a4paper, logo, copyright, nonumbering]{xiaomi}

\usepackage[numbers, sort&compress, square]{natbib}
\usepackage{dblfloatfix}
\usepackage[normalem]{ulem}
\usepackage{caption}
\usepackage{dramatist}
\usepackage{xspace}
\usepackage{pifont} 
\usepackage{multirow}
\usepackage{tcolorbox}
\usepackage{xltabular}
\usepackage{longtable}
\usepackage{hyperref}
\usepackage{wrapfig}

\usepackage{amsfonts}
\usepackage{amsmath}
\usepackage{amssymb}
\usepackage{lineno}
\usepackage{multirow}
\usepackage{adjustbox}

\usepackage[bottom]{footmisc}

\usepackage{CJKutf8}
\usepackage{setspace}
\usepackage{makecell}
\usepackage{graphicx}
\usepackage{subcaption}
\usepackage{multicol} 

\usepackage[utf8]{inputenc} 
\usepackage[T1]{fontenc}    
\usepackage{hyperref}
\usepackage{cleveref}
\usepackage{url}            
\usepackage{booktabs}       
\usepackage{amsfonts}       
\usepackage{nicefrac}       
\usepackage{microtype}      
\usepackage{xcolor}         
\usepackage{colortbl}
\usepackage{tcolorbox}
\usepackage{xspace}
\definecolor{BrickRed}{rgb}{.72,0,0}
\definecolor{darkgreen}{rgb}{0.0, 0.5, 0.0}
\definecolor{ForestGreen}{RGB}{34,139,34}
\definecolor{LakeBlue}{RGB}{0,61,153}
\definecolor{MiOrange}{RGB}{255,225,204}
\definecolor{Hex}{RGB}{225,213,231}

\hypersetup{
    colorlinks=true,
    allcolors=LakeBlue
}

\usepackage{listings}
\title{\centering TV-AudioRemover: Joint Text-Visual Guided Sound Removal with Multi-Task Hard-Mixture Curriculum}

\titlerunning{Joint Text-Visual Guided Sound Removal with Multi-Task Hard-Mixture Curriculum}

\author{
  Xinyue Guo$^*$,
  Jianxuan Yang$^*$$^\dagger$, 
  Daiguo Zhou$^*$, \linebreak
  Jiagao Hu,
  Yuxuan Chen,
  Fei Wang,
  Jian Luan
}

\institute{MiLM Plus, Xiaomi Inc.}

\begin{document}

\begin{abstract}

Visual object removal can eliminate a target from video frames, yet the target’s acoustic trace persists in the soundtrack, causing noticeable audio-visual inconsistency. Existing video inpainting models operate solely on pixels, while audio editing models, especially for the sound removal task, are typically text-driven and suffer from limited unimodal control, which is less effective than multimodal guidance that provides richer semantic grounding and temporal synchronization cues. In this paper, we present Text-Visual Guided Sound Removal (\textbf{TV-AudioRemover}), a target sound removal framework that leverages natural-language instructions and optional visually edited videos to suppress the sound of the removed visual object from the original audio mixture. To obtain high-quality training data, we design a pipeline to build a million-scale dataset of single-object audio-visual aligned samples, from which we synthesize mixture-target pairs tailored for model training. To effectively exploit visual context and faithfully follow instruction intent, we augment the model architecture with task tokens, generalizable instruction modeling, and modality-specific global guidance. We further adopt multi-task training to strengthen task-role understanding, together with a hard-mixture curriculum that employs semantically similar acoustic mixtures during fine-tuning to enhance fine-grained source discrimination.
To support evaluation, we introduce \textbf{AV-Remove-Bench}, a comprehensive audio-visual object removal benchmark, accompanied by dedicated objective metrics and an MLLM-based evaluation protocol. Experiments demonstrate that our method achieves state-of-the-art performance on both subjective and objective metrics for text-only-guided and text-visual-guided audio removal tasks. Project page: \url{https://yjx-research.github.io/TV-AudioRemover/}.

\end{abstract}

\begingroup
\renewcommand{\thefootnote}{}
\footnotetext{$^{*}$ Equal contribution. $^{\dagger}$ Corresponding author.}
\endgroup

\maketitle

\section{Introduction}
\label{sec:intro}

Object removal has become a fundamental operation in modern video editing. Recent video erasers can remove a visible object and synthesize plausible pixels for the uncovered region~\cite{miao2026rose,fu2026effecterase,liu2026understanding,hu2026ideal}. However, videos are not purely visual signals. A removed person may still speak, a deleted dog may still bark, and an erased vehicle may still retain engine noise in the soundtrack. A common workaround is to mute the entire audio track or regenerate the soundtrack, but both choices are undesirable. Muting destroys useful background ambience, speech, or music, while full regeneration often changes timing, reverberation, and source identity. A practical system should instead perform selective acoustic erasure: remove only the sound associated with the visually removed target and preserve everything else.

\begin{figure}[t]
    \centering
    \includegraphics[width=0.7\linewidth]{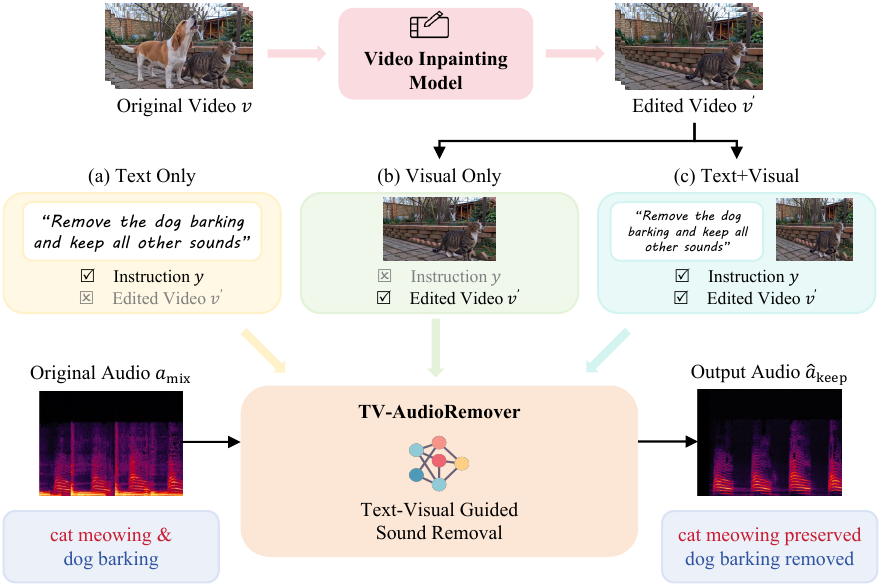}
    \caption{Overview of the Text-Visual Guided Target-Sound Removal framework TV-AudioRemover.}
    \label{fig:teaser}
\end{figure}

There are two main families of models that perform audio editing: audio-visual editing models that jointly modify video frames and sound under shared control, and audio editing models that edit only audio while taking text, reference audio, or visual queries as multimodal control signals. Although both directions have made rapid progress, they still exhibit a structural limitation for post-removal scenarios. 
For audio-visual editing models, existing strategies typically either leverage text instructions to synchronize visual-frame and audio editing~\cite{lin2026zero,xu2025schrodinger,liang2026spongebob}, or edit one modality first and then use the edited output to guide the other modality~\cite{zheng2026audio,ishii2025coherent}. However, audio-visual editing models typically center on visual manipulation. Although they support diverse editing modes, they deliver suboptimal performance on individual tasks. Furthermore, most existing audio-visual editing models are designed for human-centric general editing tasks, making them more suitable for human-speech-related editing rather than general sound effect removal. 
For audio-only editing models, the visual signal is either underused---with the model relying mainly on text and mixed audio~\cite{wang2023audit,ungersbock2026sao,tao2025mmedit}---or the task is formulated as visual-query-based sound separation~\cite{cheng2025omnisep,takahashi2026mmaudiosep,shi2025sam}, which does not transfer well to deletion-oriented settings. 

We introduce \textbf{TV-AudioRemover}, a sound removal model with multimodal control tailored for audio-visual media requiring sound elimination. As illustrated in Figure~\ref{fig:teaser}, given the original mixed audio, a textual instruction, and an optional target-object-removed video, our model aims to generate target audio that suppresses undesirable source sounds while fully preserving edit-irrelevant audio content. 
Faced with the challenges of mapping acoustic components to visual entities and distinguishing mutually exclusive extraction and removal objectives in text-visual-controlled target sound removal, we introduce improvements to data construction, model architecture, and training strategy. 
\textbf{In terms of data construction}, we first develop a dedicated pipeline to construct a million-scale, high-quality dataset of single-object audio-visual aligned samples. It combines MLLM recognition~\cite{xu2025qwen3omnitechnicalreport}, SAM Audio segmentation, and quality filtering based on CLAP~\cite{laionclap2023} and SAJ scores~\cite{wang2026samaudiojudgeunified}. Afterwards, we perform audio mixing based on the curated dataset to synthesize mixture-target pairs tailored for the target sound removal task. 
\textbf{As for model architecture}, we introduce task tokens together with generalized instruction modeling to improve instruction following. To leverage sufficient visual information without conflicting with textual features, we employ modality-specific global guidance, where dedicated global conditions are applied to each modality branch. With the above design, the model determines preservable audio content corresponding to retained visual elements. Meanwhile, textual instructions remain focused on clarifying the editing intent. 
\textbf{Regarding the training strategy}, we jointly apply two optimization schemes. On one hand, multi-task supervised learning integrates extraction, removal, and joint editing tasks. It enhances the model’s awareness of task boundaries and target sound source localization, enabling the model to differentiate audio preservation and suppression. On the other hand, we adopt a two-stage curriculum learning framework. The pre-training stage utilizes semantically distant mixtures for the model to preliminarily acquire audio elimination capability, while the fine-tuning stage leverages hard mixtures with similar semantic and acoustic features to boost fine-grained source discrimination. 
\textbf{For evaluation}, we construct AV-Remove-Bench, an audio-visual target removal benchmark with rich scene diversity and broad acoustic coverage. This benchmark consists of samples spanning public data, generated data, and recorded real-world data, covering speech, music, and sound effects. Furthermore, based on this dataset, we propose a comprehensive set of objective metrics as well as an MLLM-based evaluation protocol for the sound removal task. 
Finally, we compare our method against a variety of audio editing baselines. Experiments demonstrate that TV-AudioRemover achieves state-of-the-art performance on both subjective and objective metrics for text-only-guided and text-visual-guided audio removal tasks. 
In summary, our main contributions are as follows:
\begin{itemize}
    \item A high-quality data construction pipeline. It generates million-scale single-object audio-visual-aligned samples and synthesizes task-specific mixture-target pairs for training.
    \item A modality-decoupled audio editing architecture. With task tokens and global modality guidance, the model effectively distinguishes preservable and suppressible audio under joint visual-textual conditions.
    \item A multi-task curriculum training strategy. Combined multi-task supervision and two-stage curriculum learning enhance task understanding and fine-grained acoustic discrimination. 
    \item AV-Remove-Bench, an audio-visual target removal benchmark with rich scene diversity and broad acoustic coverage, equipped with dedicated objective metrics and an MLLM-based evaluation protocol.
\end{itemize}

\section{Related Work}
\label{sec:related}

\subsection{Video Object Removal}
Video object removal aims to erase target objects and fill the revealed region with temporally coherent content. ROSE constructs synthetic object-effect pairs and uses Diffusion Transformers to remove objects together with shadows, reflections, and illumination effects~\cite{miao2026rose}. EffectErase and UnderEraser further emphasize object-induced side effects and scene understanding~\cite{fu2026effecterase,liu2026understanding}. SVOR studies stable video object removal under imperfect real-world conditions such as mask drops and abrupt motion~\cite{hu2026ideal}. 
Existing literature demonstrates that video object removal techniques have achieved remarkable maturity. While these methods yield visually plausible videos, they leave the original audio track untouched. In practical editing scenarios, however, users expect erased objects to disappear across both modalities. Accordingly, integrating an audio editing model upon video object removal brings substantial practical value.

\subsection{Audio Editing}
Text-guided audio editing modifies an input waveform based on natural language instructions. Training-free diffusion editing methods include inversion-based approaches such as ZETA and AudioEditor, which invert the input audio into the diffusion trajectory for prompt-based editing~\cite{manor2024zero,jia2025audioeditor}, as well as inversion-free approaches such as AudioMorphix and DirectAudioEdit~\cite{liang2025audiomorphix,ge2026directaudioedit}. 
Meanwhile, training-based unified audio editing models include Audio-Omni, UNISON, and AudioWeave, which adopt shared DiT-based architectures optimized via joint multi-task training to unify audio generation and instruction-guided editing across diverse manipulation tasks~\cite{tian2026audio,li2026unison,dong2026unified}. 
Besides text-guided audio editing models, several models leverage visual conditions for audio manipulation, such as OmniSep, MMAudioSep, and SAM Audio. Nevertheless, these are sound extraction models designed for source separation driven by multimodal queries, and they are not tailored for multimodal target sound removal. 
Specifically, although multimodal extraction models such as OmniSep and MMAudioSep can accept edited videos and negative text prompts for removal tasks, they suffer from fundamental paradigm conflicts. In such models, the video query serves to extract dominant sounds corresponding to visible video content, whereas negative text prompts are used to eliminate target sounds while preserving all other audio components. This paradigm works reliably when the audio to be edited contains only two sound sources and acoustic components can be reliably mapped to visual elements. However, in scenarios where the input audio contains three or more sound sources or incorporates background audio, a modality mismatch arises: the model struggles to decide whether to match audio components to video content or to remove sounds solely according to negative text prompts. On one hand, when retained visual elements correspond to multiple sound sources, the model cannot simultaneously extract all their associated acoustic components. On the other hand, extracting audio tied to visible video content risks erroneously deleting visually unobservable background audio. In such scenarios, visual guidance becomes ineffective, and these models essentially degenerate into pure text-guided removal paradigms. Unlike these prior works, our framework trains the visual condition for retention rather than extraction in sound removal tasks.

\subsection{Audio-Visual Editing}
The first category of audio-visual editing methods modifies audio and visual content under shared instruction control. Language-guided AVEdit~\cite{liang2024language} uses text instructions to adapt paired image and audio events; Object-AVEdit~\cite{fu2025object} extends this idea to object-level manipulation, while InstructAV2AV and JAVEdit further study instruction-following joint audio-video editing at broader task levels~\cite{zheng2026instructav2av,chen2026javedit}. 
These methods do not fully exploit multimodal interactions and exhibit limited performance on specialized tasks such as audio-visual removal. Instead, they are more suited to target replacement — for example, swapping a cat for a dog while preserving the original object region and audio timing, where only the object category changes within a fixed spatial extent. Furthermore, most of these models are designed for human-centric editing tasks and lack training on and modeling of general sound effects. They tend to fail when sound elements cannot be aligned with visual content. 
The second category edits one modality first and then uses the edited result to guide the other modality. AVI-Edit uses an audio agent and audio-synchronized cues to guide instance-level video editing with mask refinement. Although such cascaded architectures perform audio and video editing separately, each module is tightly coupled and cannot be replaced. Errors introduced by the first-stage audio agent propagate to subsequent stages, and this audio agent is not optimizable. In another case, CAVE leverages edited videos to guide the audio generation process, yet it merely supports limited audio style transfer and lacks native capability for visual-object-based sound removal. 
Unlike the aforementioned approaches, TV-AudioRemover is designed as a plug-and-play downstream audio remover that accepts outputs from arbitrary visual object removal methods. This decoupled formulation facilitates reusing powerful visual removal models, isolating failure sources, and optimizing the audio module alone. It also imposes no constraints on visual conditioning and supports arbitrary input combinations: text-only, vision-only, and joint text-vision. 

In addition, open-source benchmarks for audio-visual removal remain scarce in terms of both data volume and category diversity. For instance, JAVEditBench proposed alongside JAVEdit only includes 25 audio-visual removal test samples, all of which are human-centric.

\section{Method}
\label{sec:method}

As shown in Figure~\ref{fig:teaser}, given the original mixed audio \(\mathbf{a}_{\mathrm{mix}}\), an instruction \(\mathbf{y}\), and optional edited video \(\mathbf{v}'\) obtained via a video inpainting model, TV-AudioRemover generates \(\hat{\mathbf{a}}_{\mathrm{keep}}\), where the target sound is removed while the remaining audio components are preserved. The details of the dataset construction, model architecture, training strategy, and evaluation benchmark are elaborated below.

\subsection{Data Construction}

\begin{figure}[h]
    \centering
    \includegraphics[width=1.0\linewidth]{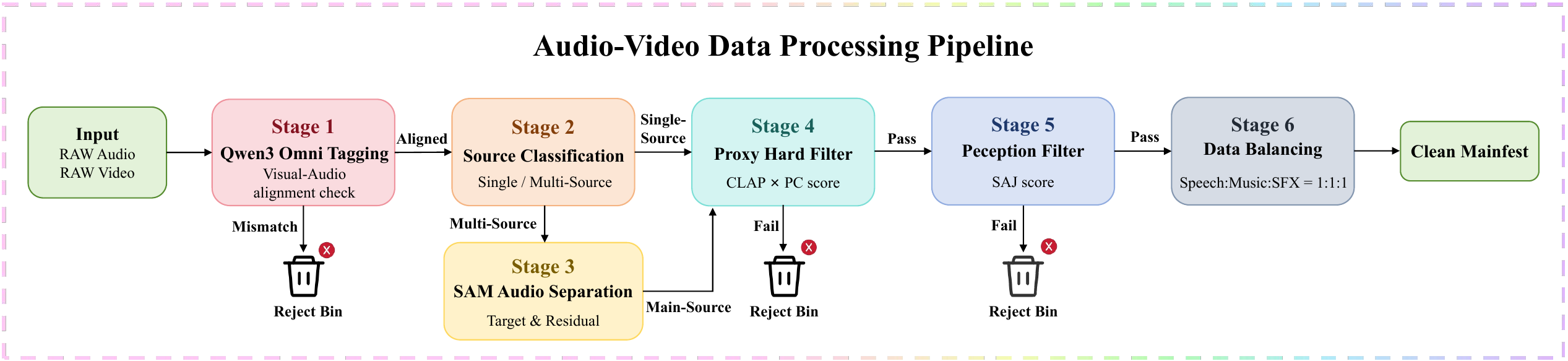}
    \caption{Data processing pipeline for constructing single-object audio-visual aligned samples.}
    \label{fig:data_pipeline}
\end{figure}

To acquire high-quality task-specific training data, we construct a million-scale dataset containing single-object audio-visual aligned samples. The data processing pipeline is shown in Figure~\ref{fig:data_pipeline}. 
The raw corpus combines audio-visual samples from video datasets, including VGGSound~\cite{chen2020vggsound}, AudioSet~\cite{gemmeke2017audio}, and ACAVCaps~\cite{niu2026acavcaps}, with audio-only samples from WavCaps\footnote{This work uses the \href{https://github.com/XinhaoMei/WavCaps}{WavCaps} dataset, which is available for research purposes only. The authors confirm that this dataset is used solely for academic and non-commercial purposes.}~\cite{mei2024wavcaps} and self-collected high-quality audio clips. We then conduct a series of annotation, validation, source separation and quality filtering operations on these raw candidates. 
Qwen3-Omni~\cite{xu2025qwen3omnitechnicalreport} is used to annotate each candidate with structured information, including scene type, source multiplicity, sound category, and the primary sound source object label; for video candidates, it further checks whether the dominant sound source is visually present as the main object. 
Candidates that fail this audio-visual correspondence check are rejected, and the remaining aligned samples are divided into single-source and multi-source subsets. Single-source samples proceed directly to quality filtering, while multi-source samples are processed by SAM Audio to extract the source corresponding to the main visual object for audio-visual samples, or the dominant sound source for audio-only samples. 
Finally, we apply a proxy hard filter based on CLAP and PC scores~\cite{tjandra2025meta}, a perceptual filter based on SAJ, and category balancing across speech, music, and sound effects to obtain the clean manifest. The Qwen3-Omni annotation protocol and filtering thresholds are detailed in Appendix \cref{sec:A}. 

Based on the curated single-object aligned samples, we synthesize task-specific mixture-target pairs for model training. For each clean audio sample $A$, we sample semantically different interference clips $B$ and optionally $C$, apply random temporal cropping and signal-to-noise ratio scaling to the interference sources, and mix them with $A$ to form both two-source mixtures for basic role supervision and three-source mixtures for more complex multi-source scenes. After obtaining the mixture and target audios, we define three edit modes:
\begin{equation}\label{eq1}
\begin{aligned}
&\text{extract:}\quad \mathbf{a}_{\mathrm{mix}}=\mathbf{A}+\mathbf{B}(+\mathbf{C}), \quad \mathbf{a}_{\mathrm{tar}}=\mathbf{A}, \\
&\text{delete:}\quad \mathbf{a}_{\mathrm{mix}}=\mathbf{A}+\mathbf{B}(+\mathbf{C}), \quad \mathbf{a}_{\mathrm{tar}}=\mathbf{A}(+\mathbf{C}), \\
&\text{both:}\quad   \mathbf{a}_{\mathrm{mix}}=\mathbf{A}+\mathbf{B}, \quad \mathbf{a}_{\mathrm{tar}}=\mathbf{A}.
\end{aligned}
\end{equation}

\begin{figure}[h]
    \centering
    \includegraphics[width=0.7\linewidth]{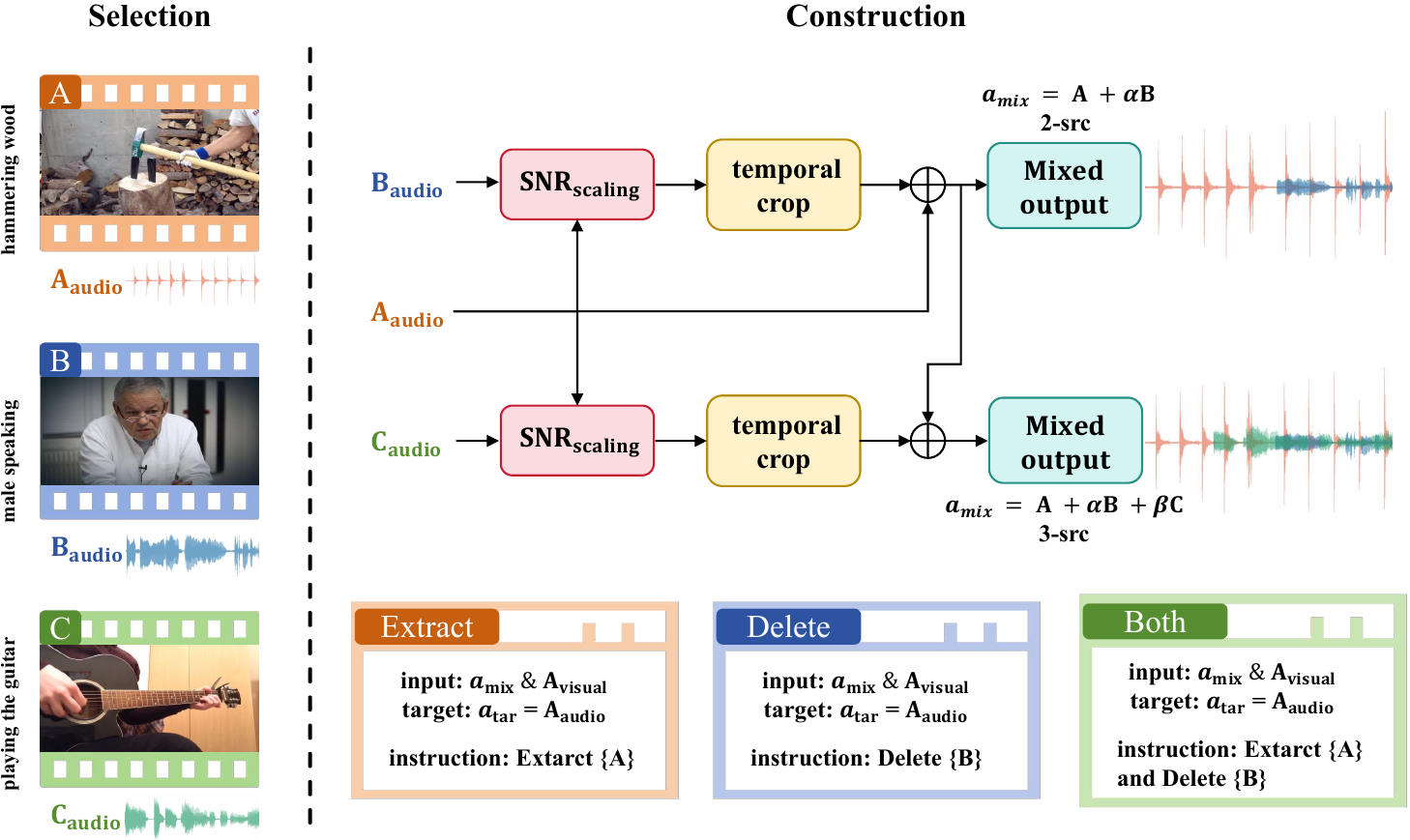}
    \caption{Selection and construction of mixture $a_{\text{mix}}$ and target $a_{\text{tar}}$ across three edit modes for synthesized training data.}
    \label{fig:training_data}
\end{figure}

For audio-visual samples, the visual condition provided to the model is typically taken from source sample $A$. Audio-only samples use empty visual features by default. 
As illustrated in Figure~\ref{fig:training_data}, the \emph{same} mixture serves as input for all three tasks; only the instruction differs, allowing one synthesized sample to provide supervision for \textsc{Extract}, \textsc{Delete}, and \textsc{Both} simultaneously. 

\subsection{Model Architecture}

\begin{figure}[t]
    \centering
    \includegraphics[width=0.9\linewidth]{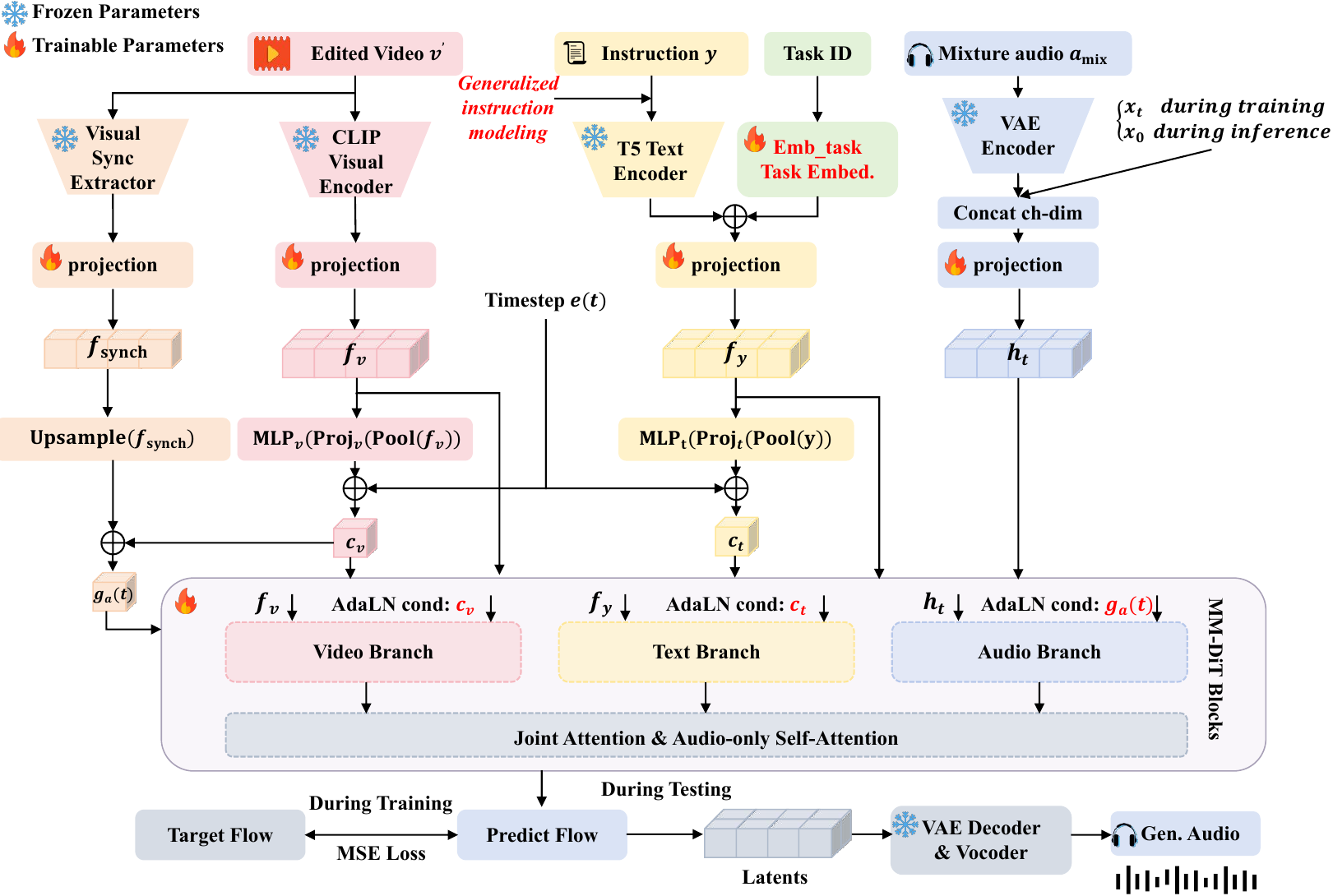}
    \caption{Architecture of the TV-AudioRemover multimodal diffusion Transformer.}
    \label{fig:architecture}
\end{figure}

TV-AudioRemover follows a latent flow-matching formulation, with the internal architecture illustrated in Figure~\ref{fig:architecture}. During training, we sample the target latent as $\mathbf{x}_1$ and the mixture latent as $\mathbf{x}_{\mathrm{cond}}$ \footnote{This work uses the Audio VAE weights from \href{https://github.com/hkchengrex/MMAudio}{MMAudio}, which are licensed under  \href{https://creativecommons.org/licenses/by-nc/4.0/}{CC BY-NC 4.0}. The authors confirm that these weights are used solely for academic and non-commercial purposes.} , normalize both latents, and construct the noisy state $\mathbf{x}_t$ between Gaussian noise $\mathbf{x}_0$ and $\mathbf{x}_1$. The current noisy latent and the mixture latent are then concatenated along the channel dimension before audio input projection:
\begin{equation}\label{eq2}
    \mathbf{x}_t = (1-t)\mathbf{x}_0 + t\mathbf{x}_1, \quad
    \mathbf{h}_t = \mathrm{Proj}_{\mathrm{audio}}\big([\mathbf{x}_t, \mathbf{x}_{\mathrm{cond}}]\big),
\end{equation}
where $[\cdot,\cdot]$ denotes channel-wise concatenation. The denoising network takes $\mathbf{h}_t$ as the audio input and predicts the flow toward $\mathbf{x}_1$, conditioned on the edited video, text instruction, and $\mathbf{x}_{\mathrm{cond}}$; at inference, the fixed $\mathbf{x}_{\mathrm{cond}}$ is concatenated with the current denoising latent at each step. 

In parallel, visual and textual features are projected into the Transformer hidden space and interact with the audio branch via MM-DiT blocks, where visual cues mark the sound sources to retain, textual instructions specify the intended operation, and the mixture audio supplies acoustic materials for reconstruction. On top of this design, the architecture introduces three key modifications: 
(1) \textbf{Task tokens.} For the three edit modes $m\in\{\mathrm{ext},\mathrm{del},\mathrm{both}\}$, we encode their task IDs as $q_m\in\{0,1,2\}$ and map them to task embeddings $\mathbf{e}_m=\mathrm{Emb}_{\mathrm{task}}(q_m)$. This embedding is added to the encoded instruction features, i.e., $\mathbf{f}_y=E_{\mathrm{txt}}(\mathbf{y})+\mathbf{e}_m$, so that the instruction condition explicitly carries the designated task role. 
(2) \textbf{Generalized instruction modeling.} We diversify instructions across task types and language forms: extraction, deletion, and combined preserve-remove commands are each described by multiple natural-language templates. This helps the model learn complementary acoustic roles, improves robustness to linguistic variation, and reduces prompt-pattern overfitting. Representative examples and the complete template set are provided in Appendix \cref{sec:B}. 
(3) \textbf{Modality-specific global guidance.} Visual and textual features are projected into separate global conditions, and the audio branch is modulated by the visual condition together with diffusion-time and synchronization cues:
\begin{equation}\label{eq3}
\begin{aligned}
&\mathbf{c}_v = \mathbf{e}(t) + \mathrm{MLP}_v\big(\mathrm{Proj}_v\big(\mathrm{Pool}(\mathbf{f}_v)\big)\big), \\
&\mathbf{c}_t = \mathbf{e}(t) + \mathrm{MLP}_t\big(\mathrm{Proj}_t\big(\mathrm{Pool}(\mathbf{f}_y)\big)\big), \\
&\mathbf{g}_a(t) = \mathbf{c}_v + \mathrm{Up}(\mathbf{f}_{\mathrm{synch}}).
\end{aligned}
\end{equation}
where $\mathbf{e}(t)$ is the diffusion-time embedding, $\mathbf{f}_{\mathrm{synch}}$ denotes synchronization-aware visual features, and $\mathrm{Up}(\cdot)$ aligns them to the audio-latent length. The visual branch is conditioned on $\mathbf{c}_v$, the audio denoising branch is guided by $\mathbf{g}_a(t)$, and the instruction branch receives $\mathbf{c}_t$. Notably, the audio branch uses visual rather than textual global modulation, since the edited video provides a positive global cue about the desired visual-acoustic state, whereas text may specify either preservation or removal.

\subsection{Training Strategy}
The training strategy consists of two components: multi-task learning across complementary edit modes, and a two-stage hard-mixture curriculum built upon source pairs of two distinct difficulty levels.
For multi-task learning, each batch contains randomly sampled instances from source extraction, source deletion, and joint editing modes. The extraction task is not mainly introduced for standalone extraction, but to strengthen target-source localization, condition alignment between visual cues and acoustic sources, and distinction between positive preservation and negative suppression. Training only with deletion samples and instructions can both overfit to specific prompt patterns and suffer from over-removal. 
For the two-stage hard-mixture curriculum, training proceeds from easy to hard mixtures. In the pre-training stage, the interference sources $B$ and $C$ mixed with source $A$ are selected as semantically distant samples that have labels different from $A$ and yield T5-based label similarity scores below a preset threshold. During fine-tuning, we instead sample hard pairs, such as male and female voices or timbre-similar instruments, to enhance fine-grained and attribute-level discrimination. Signal-to-noise ratios are also varied so that the model handles both dominant and subtle targets. Details of sample construction and benefits of multi-task training are provided in Appendix \cref{sec:C}.

We further use modality-wise classifier-free training with mutually exclusive dropout masks: both visual and text conditions are dropped with probability 0.05, only visual conditions with probability 0.15, and only text conditions with probability 0.05; when text is dropped, the task ID is also masked. This enables a unified model to operate under visual-only, text-only, and joint visual-text conditioning. Given conditions $\mathcal{C}=\{\mathbf{f}_v,\mathbf{f}_{\mathrm{synch}},\mathbf{f}_y,\mathbf{x}_{\mathrm{cond}}\}$, the network $F_\theta$ predicts the latent velocity field from noise $\mathbf{x}_0$ to target audio latent $\mathbf{x}_1$. The flow-matching objective is
\begin{equation}\label{eq4}
\mathcal{L}_{\mathrm{fm}}=\mathbb{E}_{t,\mathbf{x}_0,\mathbf{x}_1}\left[\left\|F_\theta(\mathbf{x}_t,t,\mathcal{C})-(\mathbf{x}_1-\mathbf{x}_0)\right\|_2^2\right].
\end{equation}
At inference, classifier-free guidance combines conditional predictions with an unconditional branch that masks visual and text conditions but retains $\mathbf{x}_{\mathrm{cond}}$, this adjusts the editing intent while preserving access to the source-audio reconstruction prior.

\subsection{AV-Remove-Bench}

\begin{figure}[h]
    \centering
    \includegraphics[width=0.9\linewidth]{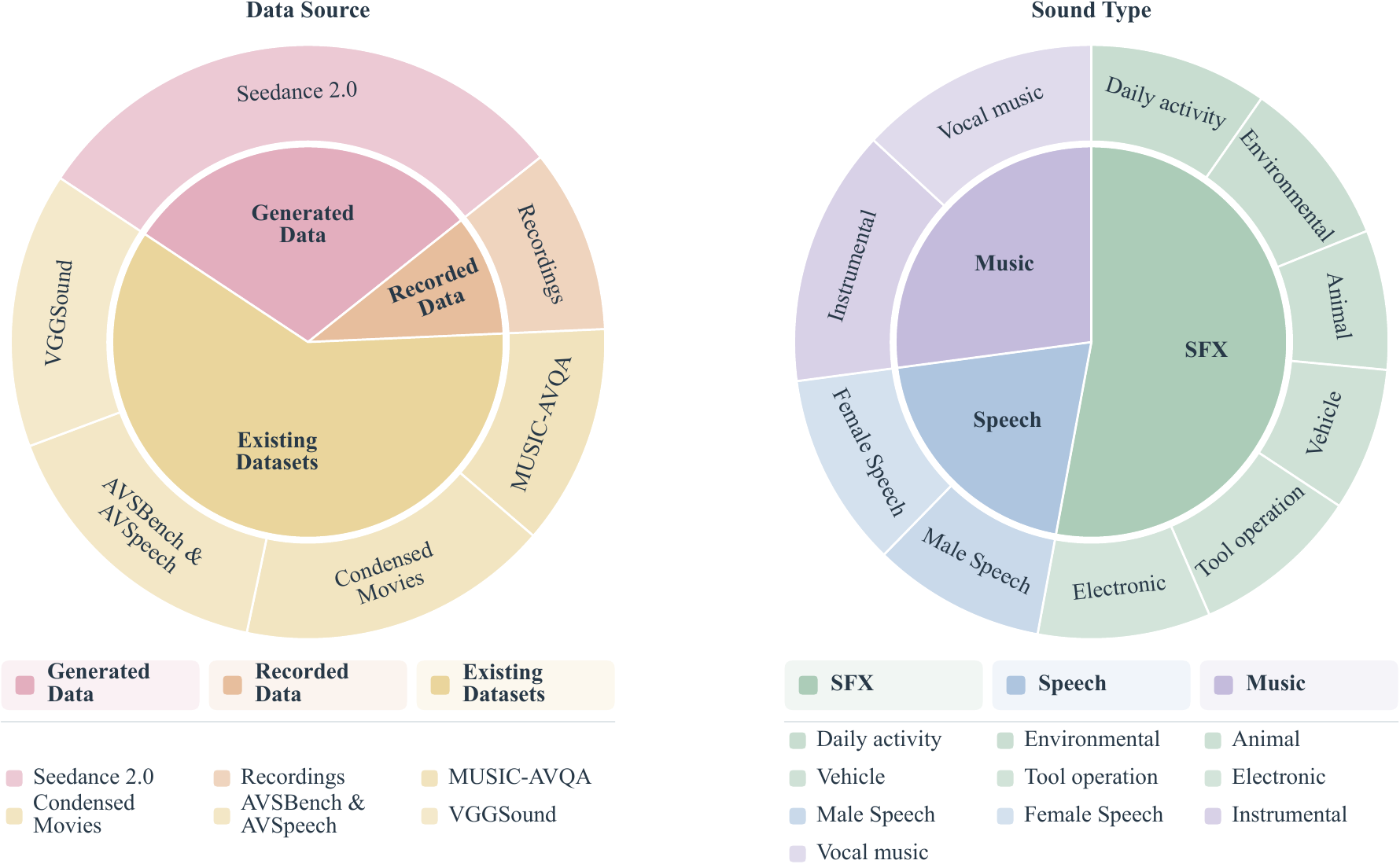}
    \caption{Data distribution of AV-Remove-Bench.}
    \label{fig:av-remove-bench}
\end{figure}

To support evaluation, we present AV-Remove-Bench, a comprehensive audio-visual object removal benchmark containing 77 samples. All samples are audio-visual clips of about six seconds, covering speech, music, and sound effects from three sources: public datasets, including MUSIC-AVQA\footnote{This work uses the \href{https://gewu-lab.github.io/MUSIC-AVQA/}{MUSIC-AVQA} dataset, which is licensed under  \href{https://creativecommons.org/licenses/by-nc/4.0/}{CC BY-NC 4.0}. The authors confirm that the aforementioned dataset in this work is used only for academic and non-commercial purposes.} ~\cite{liu2024tackling} for music, AVSpeech~\cite{ephrat2018looking} for speech, and Condensed Movies~\cite{bain2020condensed}, AVSBench\footnote{This work uses the \href{https://opennlplab.github.io/AVSBench/}{AVSBench} dataset, which is licensed under  \href{https://creativecommons.org/licenses/by-nc/4.0/}{CC BY-NC 4.0}. The authors confirm that the aforementioned dataset in this work is used only for academic and non-commercial purposes.}~\cite{zhou2022audio}, and VGGSound-test for sound effects; Seedance 2.0-generated~\cite{seedance2026seedance} videos with prompts spanning the same three audio categories; and real recordings of sound-effect-centric daily scenes. Each sample falls into either a multi-sound-emitting object setting or a foreground-target-plus-stable-background setting. For this benchmark, the task is to remove a specified sounding object from both the visual and audio streams. Further details on benchmark design are provided in Appendix \cref{sec:D}, and the data distribution is visualized in Figure~\ref{fig:av-remove-bench}. 

To thoroughly evaluate sound removal performance, we adopt two newly proposed objective metrics, Target Source Suppression Ratio (TSSR) and Preserved Source Fidelity (PSF), alongside established evaluation metrics that are described in the experimental section, to assess both target-source suppression and non-target sound preservation. 
TSSR and PSF are computed using an AST audio classifier~\cite{gong2021ast}, which predicts event-level confidence scores for the queried source categories from both the original and edited audio. Let $P_{A,\mathrm{r}}$ and $P_{B,\mathrm{r}}$ denote the confidence scores predicted for the target sound to be removed in the original audio $A$ and edited audio $B$, and let $P_{A,\mathrm{k}}$ and $P_{B,\mathrm{k}}$ denote the confidence scores predicted for the source to retain. TSSR measures the relative drop in confidence of the removed source:
\begin{equation}\label{eq5}
    \mathrm{TSSR}=10\times\frac{P_{A,\mathrm{r}}-P_{B,\mathrm{r}}}{P_{A,\mathrm{r}}}, \quad P_{A,\mathrm{r}}>\epsilon.
\end{equation}
When $P_{A,\mathrm{r}}\le\epsilon$, the classifier indicates that the original audio barely contains the target source, rendering TSSR undefined. A larger TSSR signifies stronger suppression. To quantify the preservation quality of the source to retain, PSF penalizes drops in confidence but allows a bounded bonus when editing improves the source to retain:
\begin{equation}\label{eq6}
    \mathrm{PSF}=\begin{cases}
    10-10\times(P_{A,\mathrm{k}}-P_{B,\mathrm{k}}), & P_{B,\mathrm{k}}<P_{A,\mathrm{k}},\\
    10+10\times\min(P_{B,\mathrm{k}}-P_{A,\mathrm{k}},\gamma), & P_{B,\mathrm{k}}\ge P_{A,\mathrm{k}},
    \end{cases}
\end{equation}
where $\gamma=0.5$ limits score drift. A higher PSF indicates better preservation of the non-target source.

In addition, we use Gemini 2.5 Pro~\cite{comanici2025gemini} as a multimodal judge to approximate human subjective perception. The detailed Gemini scoring criteria are provided in Appendix \cref{sec:E}.

\section{Experiments}
\label{sec:exp}

\subsection{Experimental Setup}
\textbf{Implementation details.} 
Training uses the curated single-object aligned corpus described in the Data Construction section. Each training clip is 8 seconds long. The pre-training split contains about 1.11M general clips, including 602.9K audio-visual clips and 511.2K audio-only clips; the fine-tuning split contains 5.65K hard clips with an approximately 1:1 ratio of audio-visual to audio-only data. We synthesize mixture-target pairs online rather than storing fixed mixtures, and train \textsc{Extract}, \textsc{Delete}, and \textsc{Both} modes with probabilities $0.4$, $0.4$, and $0.2$, respectively. 
We use a 44 kHz audio sampling rate and perform flow-matching inference with Euler sampling over 25 denoising steps. Optimization employs AdamW with weight decay $10^{-6}$ and gradient clipping at 1.0; pre-training runs for 350K iterations with a learning rate of $10^{-4}$ and step-based learning rate decay, and fine-tuning runs for 80K iterations from the pre-trained checkpoint with a learning rate of $2\times10^{-6}$.

\textbf{Evaluation datasets and Baselines.} 
The evaluation datasets consist of two parts: a removal subset of 50 samples from MMAE~\cite{ma2026mmae}, an open-source text-guided audio editing benchmark, and our proposed text-visual-guided sound removal benchmark AV-Remove-Bench. 
We compare against three groups of systems. The first group consists of joint audio-visual editing models AVI-Edit and InstructAV2AV, used to evaluate sound removal under an end-to-end audio-visual setting. The second group consists of pure text-guided audio editing models that do not use visual conditions, including ZETA, Audio-Omni, and UNISON; together, they cover inversion-based, instruction-guided, and LLM-guided audio editing paradigms. The third group of baselines includes SAM Audio and MMAudioSep, both video-and-text-controlled audio extraction models closely related to our task. For evaluation of SAM Audio, we reformulate removal as extracting the sources to retain for multi-sounding-object cases; for foreground-target-plus-background cases, we reformulate it as extracting the target source to be removed and adopt the residual branch for output. For fair comparison, we set the number of reranking candidates to 1 during SAM Audio inference. For MMAudioSep, we use negative prompts to guide sound removal. 
We use the best official versions of these open-source SOTA models for inference under identical experimental conditions. For MMAudioSep, since official model weights have not been publicly released, we carefully reproduce the model following its official code and data before conducting evaluations. 
Furthermore, we conduct comprehensive comparisons against commercial audio-visual editing models, including Seedance 2.0, Seedance 2.5, and MiniMax H3, and demonstrate the superiority of our method. Complete experimental results are available in Appendix \cref{sec:F}. 
Since TV-AudioRemover takes the edited video as input, we also evaluate how upstream visual removal affects the final result by pairing TV-AudioRemover with EffectErase, ROSE, UnderEraser, and SVOR. We adopt the outputs from SVOR, the best-performing visual-removal model, as the visual input to our model during the text-visual-guided audio editing evaluation.

\textbf{Metrics.} 
For the removal subset of MMAE, we adopt its official evaluation script and report three metrics: Instruction Following, Consistency, and Exact Match. MMAE adopts a rubric-based evaluation pipeline with Qwen3-Omni serving as the evaluator. Instruction Following quantifies adherence to editing instructions, Consistency assesses the preservation of instruction-irrelevant audio components, and Exact Match requires all rubric criteria to be satisfied for a passing sample. All metrics are reported on a 0–100 percentage scale. 
For AV-Remove-Bench, reference-based separation metrics such as SDR are not applicable because no ground-truth target-removed reference audio is available for evaluation samples. We therefore evaluate sound removal with both model-based metrics and MLLM-based evaluation. For model-based metrics, IS measures audio quality and diversity using a pre-trained audio classifier PANNs~\cite{kong2020panns}; SAJ Overall measures perceptual separation quality, while IB-AV and DeSync measure audio-visual consistency, where IB-AV evaluates semantic audio-visual alignment with ImageBind~\cite{girdhar2023imagebind} and DeSync estimates temporal offset with Synchformer; TSSR and PSF are removal-specific metrics proposed in the AV-Remove-Bench section. For MLLM-based metrics, we employ Instruction Compliance$_a$ and Fidelity$_a$, which correspond to scores generated by Gemini 2.5 Pro for edited audio across distinct perspectives. Detailed computation protocols are provided in Appendix~\ref{sec:G}. Furthermore, we report the number of parameters and average runtime for each model. For ZETA, we adopt AudioLDM2-Large as its backbone when counting parameters. For Audio-Omni, only the parameters of its trainable DiT generation backbone are counted. 
We further conduct human listening evaluation on AV-Remove-Bench. Scores are collected from 10 raters with professional audio-editing experience, who are blinded to model identities. Raters score each output on a three-level scale $\{0,0.5,1\}$ along four dimensions: target-removal completeness, background preservation, temporal naturalness, and overall listening quality. The first dimension checks whether the unwanted sound is still audible; the second checks whether non-target sources are weakened or accidentally removed; the third assesses discontinuities, unstable loudness, and unnatural transitions; and the last evaluates overall audio quality and whether the output retains meaningful audio content.

\subsection{Objective Results}

\begin{figure}[t]
    \centering
    \includegraphics[width=\linewidth]{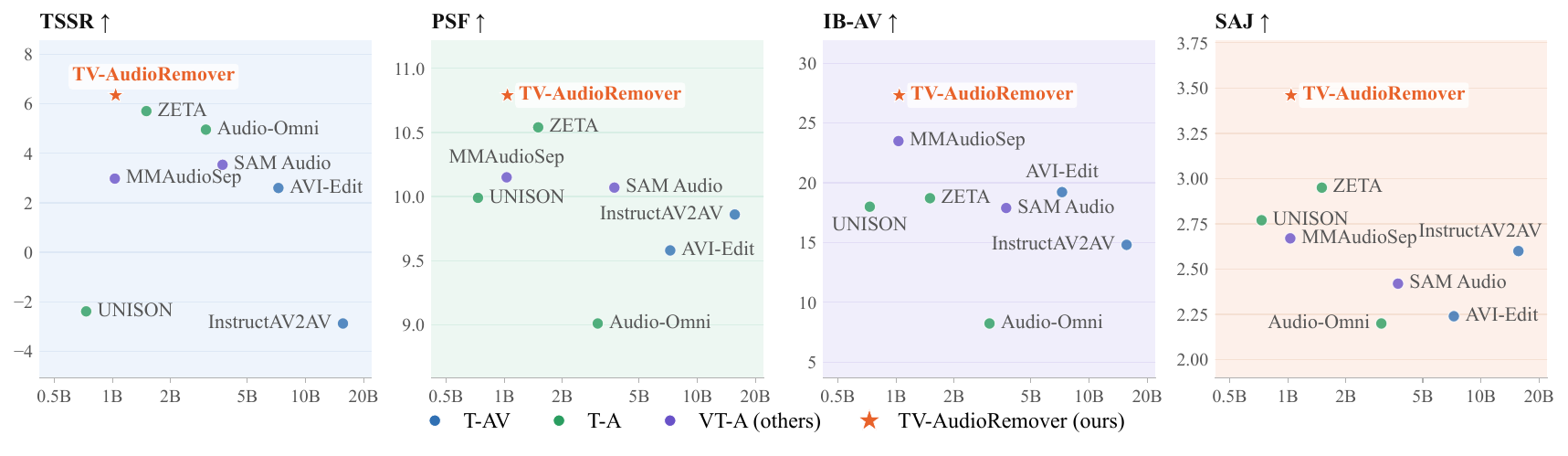}
    \caption{Comparison of key metrics and model scale on AV-Remove-Bench. Each scatter point represents one method; colors denote input-modality groups (T-AV: blue, T-A: green, VT-A: violet, ours: orange star). The x-axis is log-scaled by parameter count.}
    \label{fig:key-metrics}
\end{figure}

\begin{figure}[h]
    \centering
    \includegraphics[width=0.6\linewidth]{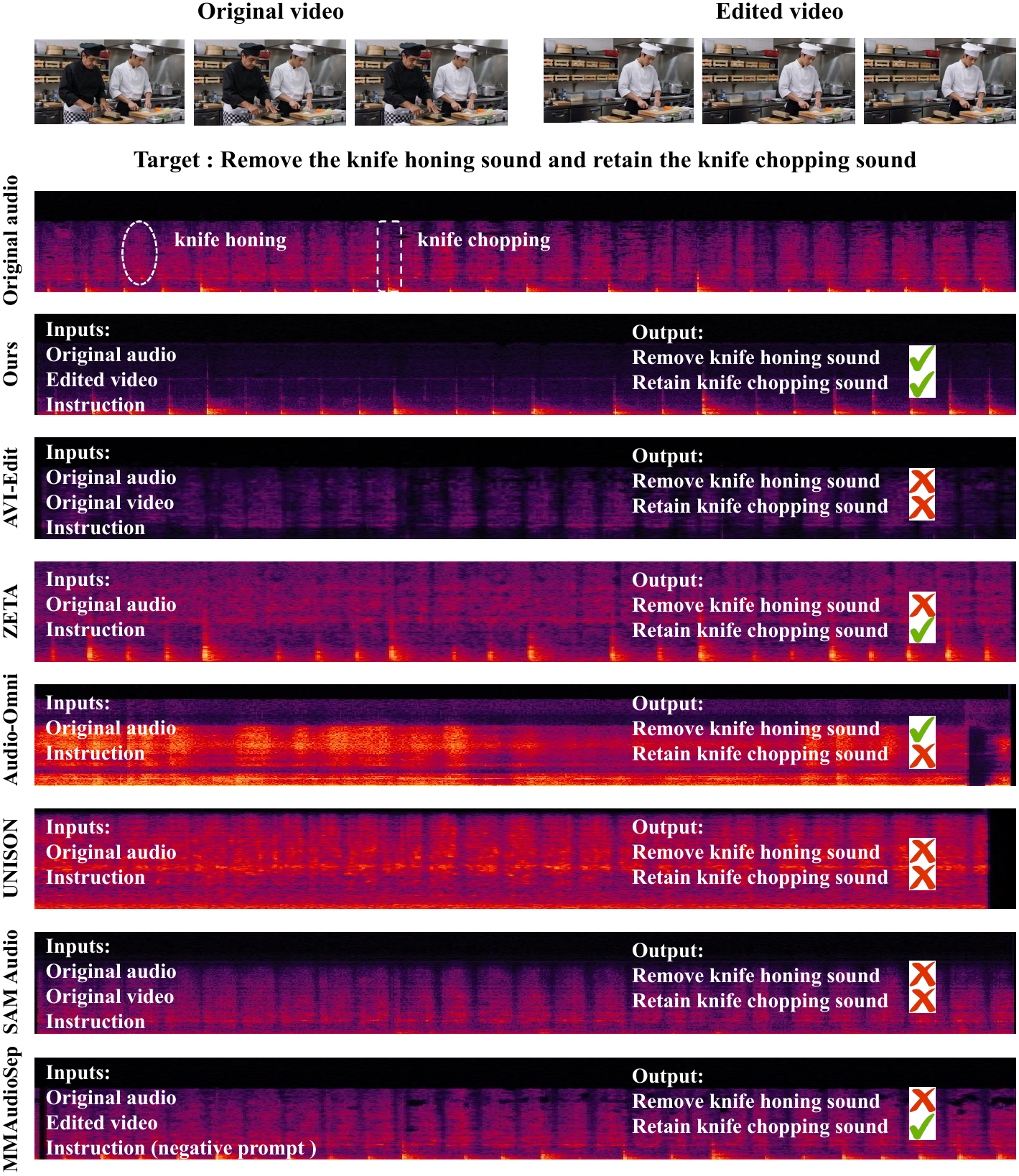}
    \caption{Qualitative comparison of different sound removal methods.}
    \label{fig:visualization}
\end{figure}

Table~\ref{tab:mmae} and Table~\ref{tab:cmp} present the objective results on the MMAE removal subset and AV-Remove-Bench. Figure~\ref{fig:key-metrics} plots the four key metrics against model parameter count across all compared methods. Qualitative visualization examples are provided in Figure~\ref{fig:visualization}. 

\begin{table}[h]
\centering
\renewcommand{\arraystretch}{0.8}
\resizebox{0.65\linewidth}{!}{
\begin{tabular}{lccc}
\toprule
Method & Instruction Following$\uparrow$ & Consistency$\uparrow$ & Exact Match$\uparrow$\\
\midrule
ZETA & 58.99 & 42.70 & 0.00\\
MMEdit & 51.70 & \textbf{82.47} & 10.42\\
Audio-Omni & 48.51 & 56.49 & 2.08\\
UNISON & \underline{72.19} & 73.77 & \underline{20.83}\\
\textbf{TV-AudioRemover} & \textbf{76.77} & \underline{82.25} & \textbf{22.92}\\
\bottomrule
\end{tabular}%
}
\caption{Objective results on the MMAE removal subset. Best results are shown in bold and second-best results are underlined.}
\label{tab:mmae}
\end{table}

Table~\ref{tab:mmae} demonstrates that our model achieves SOTA performance on the natural-language-controlled sound removal task. On this test set, we only evaluate the sound removal performance of audio-only editing models under pure text control; we do not report results of MMAudioSep and SAM-Audio on this dataset since MMAE tasks rely on natural-instruction guidance rather than object labels, and the residual components after sound removal are unknown or unspecifiable. This precludes the use of negative prompts and conversion into an extraction task. Specifically, our TV-AudioRemover achieves the best scores on Instruction Following (76.77) and Exact Match (22.92), and ranks second on Consistency (82.25), outperforming all compared baseline models. UNISON delivers the second-best results in Instruction Following and Exact Match yet attains a moderate Consistency score, indicating that it cannot well preserve non-target regions of the original audio while executing target tasks. Although MMEdit achieves the highest Consistency, this comes at the cost of compromised instruction-following capability. ZETA and Audio-Omni exhibit relatively weak performance in both instruction adherence and content preservation. In particular, ZETA fails to satisfy all editing rubrics for any sample, yielding an Exact Match score of 0. This stems from its inversion-based architecture: the model requires source audio paired with descriptions of target audio with limited variation between input and output. By contrast, MMAE only provides editing instructions and contains highly complex editing scenarios. 

\begin{table}[t]
\centering
\resizebox{\linewidth}{!}{
\renewcommand{\arraystretch}{1.0}
\begin{tabular}{llcccccccccc}
\toprule
Group & Method & \multicolumn{6}{c}{Audio metrics} & \multicolumn{2}{c}{Audio‑visual metrics} & \multicolumn{2}{c}{Efficiency}\\
\cmidrule(lr){3-8} \cmidrule(lr){9-10} \cmidrule(lr){11-12}
& & IS $\uparrow$ & SAJ Overall $\uparrow$ & TSSR $\uparrow$ & PSF $\uparrow$ & Instr.\ Comp.$_a$$\uparrow$ & Fidelity$_a$$\uparrow$ & IB‑AV $\uparrow$ & DeSync $\downarrow$ & Params & Times (s) \\
\midrule
\multirow{2}{*}{T-AV}
& AVI‑Edit & 3.16 & 2.24 & 2.60 & 9.58 & 3.56 & 2.57 & \underline{19.22} & 0.77 & 7.24B & 255.89\\
& InstructAV2AV & 3.45 & 2.60 & -2.88 & 9.86 & 2.65 & 3.04 & 14.81 & 0.76 & 15.63B & 299.55\\
\midrule
\multirow{3}{*}{T-A}
& ZETA & 3.27 & \underline{2.95} & \underline{5.71} & \underline{10.54} & 3.69 & 3.22 & 18.71 & \underline{0.75} & 1.5B & 86.98\\
& Audio‑Omni & 2.15 & 2.20 & 4.96 & 9.01 & \underline{4.39} & 1.68 & 8.23 & 1.13 & 3.05B & 6.18\\
& UNISON & \textbf{3.63} & 2.77 & -2.39 & 9.99 & 4.27 & 3.44 & 18.00 & 0.78 & 731.71M & 5.82\\
\midrule
\multirow{2}{*}{VT-A} & SAM Audio & 3.23 & 2.42 & 3.54 & 10.07 & 3.70 & 3.90 & 17.89 & 0.93 & 3.72B & 1.94\\
& MMAudioSep & 3.48 & 2.67 & 2.98 & 10.15 & 3.41 & \underline{3.94} & 23.49 & 0.97 & 1.03B & 2.03\\
\midrule
VT-A (Ours) & \textbf{TV-AudioRemover} & \underline{3.55} & \textbf{3.46} & \textbf{6.35} & \textbf{10.79} & \textbf{4.79} & \textbf{3.96} & \textbf{27.32} & \textbf{0.60} & 1.04B & 2.07\\
\bottomrule
\end{tabular}%
}
\caption{Objective evaluation on the AV-Remove-Bench. Best results are shown in bold and second-best results are underlined.}
\label{tab:cmp}
\end{table}

Next, the results in Table~\ref{tab:cmp} --- our core table presenting objective results on AV-Remove-Bench --- are analyzed in detail below. For audio-only editing methods, results for IB-AV and DeSync are obtained by pairing their sound-removal outputs with SVOR’s edited videos. SVOR, validated as the best-performing visual remover in Appendix~\ref{sec:H}, provides visual input to our model and MMAudioSep. For SAM Audio, given raw audio-visual clips and a target-object mask, SAM 2 performs spatio-temporal segmentation of the target object across all visual frames, and the resulting masks are fed into SAM Audio as visual prompts. For PSF, we only compute metrics for cases where the original audio contains two sounding objects, such that the source to be preserved can be described by a single label. 
Audio-only editing models show different strengths but exhibit unbalanced performance on sound removal tasks. Although UNISON achieves the highest IS metric, which reflects general audio quality, it yields low TSSR and PSF, indicating limited suppression of target sound sources and limited preservation of non-target sound sources. This occurs because the model retains a large portion of the original input mixture, leading to an inflated IS score. Audio-Omni obtains moderate instruction compliance, but its low Fidelity$_a$ reveals that the model incurs substantial damage to the preserved audio when performing target sound removal. ZETA achieves competitive scores on several metrics, as we perform instruction adaptation for the model and the source and target audio of test samples can be clearly specified. Nevertheless, its performance still has notable limitations. 
SAM Audio and MMAudioSep exhibit moderate overall performance, with balanced results across metrics and no outstanding strengths or weaknesses. Comprehensive comparisons against our model reveal that extraction-oriented models are not directly optimized for target sound suppression and residual source preservation. 
TV-AudioRemover achieves the best TSSR, PSF, Instruction Compliance$_a$, and Fidelity$_a$, demonstrating stronger target suppression while preserving non-target sources. It also obtains the best IB-AV and DeSync scores, indicating that the edited audio is more semantically and temporally consistent with the edited video and demonstrating the advantages introduced by visual input for audio-visual consistency. 
Figure~\ref{fig:key-metrics} plots the key metrics against model parameter count across all compared methods, showing that TV-AudioRemover achieves the highest scores on all metrics with only 1.04B parameters---on par with the lightest baselines---indicating that the gains stem from architecture rather than scale. Representative audio-visual editing examples are presented in  Figure~\ref{fig:visualization} to qualitatively illustrate the removal quality.

\subsection{Subjective Results}

\begin{table}[h]
\centering
\resizebox{\linewidth}{!}{
\setlength{\tabcolsep}{3pt}
\begin{tabular}{lcccccc}
\toprule
\multicolumn{7}{l}{\textbf{Human subjective scores}}\\
Method & Target Remove Comp. $\uparrow$ & Background Pres. $\uparrow$ & Temporal Nat. $\uparrow$ & Overall Qual. $\uparrow$ & AC $\uparrow$ & ASR $\uparrow$ \\
\midrule
AVI-Edit & 0.51 & 0.32 & 0.35 & 0.35 & 39.37 & 21.43\% \\
InstructAV2AV & 0.38 & 0.43 & 0.28 & 0.31 & 37.20 & 18.86\% \\
ZETA & 0.42 & 0.63 & 0.55 & 0.55 & 53.36 & 36.69\% \\
Audio-Omni & 0.48 & 0.24 & 0.20 & 0.19 & 30.75 & 11.69\% \\
UNISON & 0.65 & 0.60 & 0.60 & 0.62 & 62.00 & 49.35\% \\
SAM Audio & \underline{0.86} & 0.83 & 0.89 & \underline{0.88} & \underline{85.72} & \underline{81.58\%} \\
MMAudioSep & 0.75 & \underline{0.88} & \underline{0.92} & 0.81 & 83.00 & 80.74\% \\
\textbf{TV-AudioRemover} & \textbf{0.98} & \textbf{0.89} & \textbf{0.94} & \textbf{0.95} & \textbf{93.80} & \textbf{97.40\%} \\
\bottomrule
\end{tabular}
}
\caption{Human subjective scores on the AV-Remove-Bench.}
\label{tab:user_study_human}
\end{table}

Table~\ref{tab:user_study_human} presents human listening scores.
Here, Target Remove Comp. (TRC), Background Pres. (BP), Temporal Nat. (TN), and Overall Qual. (OQ) denote target removal completeness, background preservation, temporal naturalness, and overall quality, respectively. AC denotes the audio composite score, computed as $(0.35\times$TRC $+0.35\times$BP $+0.15\times$TN $+0.15\times$OQ$)\times 100$, and ASR denotes the audio success rate, i.e., the fraction of samples whose four human scores are all non-zero. 
In the human evaluation, TV-AudioRemover ranks first on all reported metrics. Against SAM Audio, a strong video-text baseline, TV-AudioRemover gains 13.8\% on TRC, 7.3\% on BP, 6.0\% on TN, 7.5\% on OQ, 9.4\% on AC, and 19.4\% on ASR, highlighting the benefit of task-specific tuning for sound removal. Compared with strong text-guided audio editing baselines such as UNISON, TV-AudioRemover further improves all metrics, showing that visual conditioning enhances source identification and preserves scene-consistent non-target sounds. 
Notably, ZETA achieves lower subjective scores relative to its objective metrics. This discrepancy may stem from the different evaluation dimensions: objective metrics focus on low-level waveform matching achieved by ZETA’s inversion process, while human listeners assess the edited audio from the perspective of high-level auditory semantics, where the output fails to fulfill the target sound removal and non-target source retention requirements.

\subsection{Ablation}

\begin{table}[h]
\centering
\setlength{\tabcolsep}{3pt}
\renewcommand{\arraystretch}{0.8}
\begin{tabular}{lccccc}
\toprule
Setting & SAJ Overall $\uparrow$ & TSSR $\uparrow$ & PSF $\uparrow$ & IB-AV $\uparrow$ & DeSync $\downarrow$ \\
\midrule
\multicolumn{6}{l}{\textbf{Training strategy}} \\
Single-stage & 3.22 & 4.82 & 10.51 & 25.17 & 0.62 \\
Two-stage & \textbf{3.46} & \textbf{6.35} & \textbf{10.79} & \textbf{27.32} & \textbf{0.60} \\
\midrule
\multicolumn{6}{l}{\textbf{Training task}} \\
Single-task & 3.11 & 5.16 & 10.43 & 26.90 & 0.69 \\
Multi-task & \textbf{3.46} & \textbf{6.35} & \textbf{10.79} & \textbf{27.32} & \textbf{0.60} \\
\midrule
\multicolumn{6}{l}{\textbf{Input modalities}} \\
Text only & \textbf{3.46} & 6.31 & 10.73 & 25.02 & 0.63 \\
Visual only & 2.97 & -3.64 & 10.27 & 26.72 & 0.67 \\
Visual + Text & \textbf{3.46} & \textbf{6.35} & \textbf{10.79} & \textbf{27.32} & \textbf{0.60} \\
\midrule
\multicolumn{6}{l}{\textbf{Upstream visual removers}} \\
ROSE & 3.29 & 6.27 & 10.72 & 26.07 & 0.77 \\
EffectErase & 3.22 & 5.52 & 10.65 & 24.21 & 0.89 \\
UnderEraser & 3.23 & 5.84 & 10.71 & 26.12 & 0.75 \\
SVOR & \textbf{3.46} & \textbf{6.35} & \textbf{10.79} & \textbf{27.32} & \textbf{0.60} \\
\bottomrule
\end{tabular}
\caption{Ablation studies of TV-AudioRemover including training strategy, input modalities and upstream visual removers.}
\label{tab:all_ablation}
\end{table}

\begin{figure}[h]
    \centering
    \includegraphics[width=1.0\linewidth]{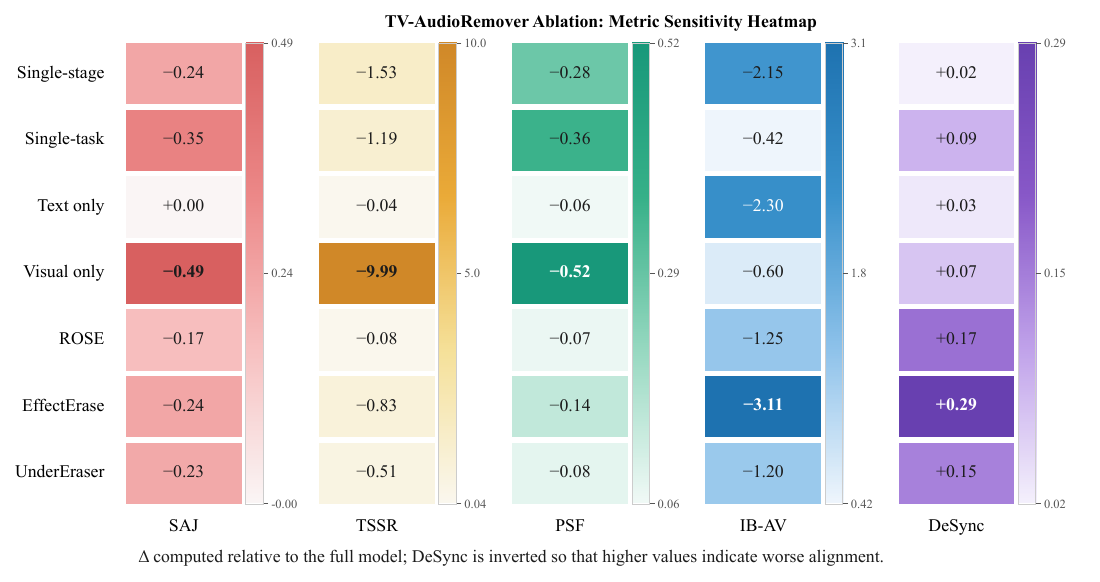}
    \caption{Performance variation $\Delta$ of TV-AudioRemover under different ablation settings, measured relative to the full model. $\Delta$ is computed relative to the full model. For SAJ, TSSR, PSF and IB-AV, negative $\Delta$ indicates performance degradation. For DeSync, positive $\Delta$ means worse alignment. Darker red represents larger performance drops, and blue denotes mild degradation.}
    \label{fig:ablation_heatmap}
\end{figure}

Tables~\ref{tab:all_ablation} summarize the ablation results on four aspects. 
First, the \textbf{training-strategy ablation} shows that two-stage training is crucial for sound removal quality. Compared with single-stage training, the two-stage variant improves all metrics and is especially beneficial for target suppression. This is likely because pre-training establishes general sound removal capability, while fine-tuning sharpens fine-grained source discrimination. 
Second, the \textbf{multi-task training strategy} outperforms single-task training across all evaluation metrics, which demonstrates that joint optimization over multiple related tasks effectively boosts the overall performance of TV-AudioRemover.
Third, the \textbf{input-modality ablation} shows that text remains a strong control signal, while visual information provides complementary grounding. Text-only conditioning yields the same SAJ Overall score as the full model and only slightly lower TSSR and PSF, suggesting that linguistic instructions already specify most removal targets. However, when the model struggles to interpret the instructions, the visual stream supplies additional semantic and temporal cues, as evidenced by the higher IB-AV and lower DeSync under joint visual-text conditioning. Under visual-only conditioning, TSSR decreases substantially. The model still succeeds on a considerable subset of samples, indicating that visual information can guide source removal but cannot deliver precise editing without textual task intent. Overall, joint visual-text conditioning achieves the best performance across all metrics. 
Fourth, the \textbf{upstream-visual-remover ablation} further reveals that the quality of edited videos affects model performance when visual conditions are provided as input. In particular, when the target is not fully eliminated in the visual domain (e.g., EffectErase and UnderEraser), the model’s TSSR metric degrades substantially. When SVOR serves as the upstream visual-removal module, TV-AudioRemover attains optimal results on all metrics. These trends indicate that higher-quality visual removal produces cleaner, temporally consistent visual conditions to facilitate sound removal. Therefore, when visual conditions are adopted as input, it is necessary to verify the accuracy of edited visuals so that they bring positive enhancement rather than performance degradation. Without reliable visual cues, the model can fall back to text-only guidance. 
All experiments above demonstrate that our multimodal-guided sound removal model, on the one hand, is capable of performing matched sound removal for edited videos to remedy modality deficiency. On the other hand, compared with existing audio editing models, our method achieves SOTA performance on text-only tasks. Incorporating the visual modality further boosts model performance and improves audio-video consistency, a capability that many existing audio editing models do not possess. Figure~\ref{fig:ablation_heatmap} further visualizes the performance variation across all ablation configurations.

\section{Conclusion}
\label{sec:conclusion}
In this paper, we propose TV-AudioRemover, a target sound removal system tailored for audio-visual media requiring sound elimination. To support this task, we construct a million-scale corpus of single-object audio-visual aligned samples, and further introduce AV-Remove-Bench for evaluation. Built on an MM-DiT-based audio editing architecture with task tokens and global modality guidance, and trained with a multi-task two-stage curriculum, TV-AudioRemover better distinguishes preservable and suppressible audio components under joint visual-textual conditions. Experiments show that the proposed system achieves state-of-the-art performance on both objective and subjective metrics against audio-visual joint editing models and audio-only editing models, demonstrating the effectiveness of our data construction, model design, and training strategy for target sound removal.

\bibliography{ref}

@article{miao2026rose,
  title={Rose: Remove objects with side effects in videos},
  author={Miao, Chenxuan and Feng, Yutong and Zeng, Jianshu and Gao, Zixiang and Liu, Hantang and Yan, Yunfeng and Qi, Donglian and Chen, Xi and Wang, Bin and Zhao, Hengshuang},
  journal={Advances in Neural Information Processing Systems},
  volume={38},
  pages={149140--149162},
  year={2026}
}

@inproceedings{fu2026effectErase,
  title={{EffectErase}: Joint Video Object Removal and Insertion for High-Quality Effect Erasing},
  author={Fu, Yang and Zheng, Yike and Dai, Ziyun and Ding, Henghui},
  booktitle={Proceedings of the IEEE/CVF Conference on Computer Vision and Pattern Recognition (CVPR)},
  year={2026}
}

@article{liu2026understanding,
  title={From Understanding to Erasing: Towards Complete and Stable Video Object Removal},
  author={Liu, Dingming and Wang, Wenjing and Li, Chen and Lyu, Jing},
  journal={arXiv preprint arXiv:2604.01693},
  year={2026}
}

@article{hu2026ideal,
  title={From Ideal to Real: Stable Video Object Removal under Imperfect Conditions},
  author={Hu, Jiagao and Chen, Yuxuan and Li, Fuhao and Wang, Zepeng and Wang, Fei and Zhou, Daiguo and Luan, Jian},
  journal={arXiv preprint arXiv:2603.09283},
  year={2026}
}

@inproceedings{liang2024language,
  title={Language-guided joint audio-visual editing via one-shot adaptation},
  author={Liang, Susan and Huang, Chao and Tian, Yapeng and Kumar, Anurag and Xu, Chenliang},
  booktitle={Proceedings of the Asian Conference on Computer Vision},
  pages={1011--1027},
  year={2024}
}

@inproceedings{lin2026zero,
  title={Zero-shot audio-visual editing via cross-modal delta denoising},
  author={Lin, Yan-Bo and Lin, Kevin and Yang, Zhengyuan and Li, Linjie and Wang, Jianfeng and Lin, Chung-Ching and Wang, Xiaofei and Bertasius, Gedas and Wang, Lijuan},
  booktitle={2026 IEEE/CVF Winter Conference on Applications of Computer Vision (WACV)},
  pages={7344--7354},
  year={2026},
  organization={IEEE}
}

@article{fu2025object,
  title={Object-AVEdit: An Object-level Audio-Visual Editing Model},
  author={Fu, Youquan and Si, Ruiyang and Wang, Hongfa and Zhou, Dongzhan and Sun, Jiacheng and Luo, Ping and Hu, Di and Zhang, Hongyuan and Li, Xuelong},
  journal={arXiv preprint arXiv:2510.00050},
  year={2025}
}

@article{ishii2025coherent,
  title={Coherent Audio-Visual Editing via Conditional Audio Generation Following Video Edits},
  author={Ishii, Masato and Hayakawa, Akio and Shibuya, Takashi and Mitsufuji, Yuki},
  journal={arXiv preprint arXiv:2512.07209},
  year={2025}
}

@article{xu2025schrodinger,
  title={Schrodinger Audio-Visual Editor: Object-Level Audiovisual Removal},
  author={Xu, Weihan and Cheng, Kan Jen and Saito, Koichi and Mirza, Muhammad Jehanzeb and Li, Tingle and Liu, Yisi and Liu, Alexander H and Wang, Liming and Ishii, Masato and Shibuya, Takashi and others},
  journal={arXiv preprint arXiv:2512.12875},
  year={2025}
}

@inproceedings{zheng2026audio,
  title={AVI-Edit: Audio-sync Video Instance Editing with Granularity-Aware Mask Refiner},
  author={Zheng, Haojie and Weng, Shuchen and Liu, Jingqi and Yang, Siqi and Shi, Boxin and Wang, Xinlong},
  booktitle={Proceedings of the IEEE/CVF Conference on Computer Vision and Pattern Recognition},
  pages={23150--23160},
  year={2026}
}

@article{zheng2026instructav2av,
  title={InstructAV2AV: Instruction-Guided Audio-Video Joint Editing},
  author={Zheng, Haojie and Yang, Yixin and Yang, Siqi and Weng, Shuchen and Shi, Boxin},
  journal={arXiv preprint arXiv:2605.18467},
  year={2026}
}

@article{liang2026spongebob,
  title={SpongeBob: Sync-Aware Harmonious Audio-Visual Generative Editing},
  author={Liang, Sen and Wang, Cong and Guan, Fengbin and Yu, Zhentao and Lu, Yiting and Wang, Yuanzhi and Zhou, Yuan and Li, Xin and Chen, Zhibo},
  journal={arXiv preprint arXiv:2605.25193},
  year={2026}
}

@article{chen2026javedit,
  title={JAVEDIT: Joint Audio-Visual Instruction-Guided Video Editing with Agentic Data Curation},
  author={Chen, Yinan and Lin, Chuming and Chen, Zhennan and Zeng, Yuxiang and Zhu, Junwei and Bi, Yali and Huang, Xijie and Xu, Chengming and Luo, Donghao and Xue, Zhucun and others},
  journal={arXiv preprint arXiv:2606.03168},
  year={2026}
}

@article{wang2023audit,
  title={Audit: Audio editing by following instructions with latent diffusion models},
  author={Wang, Yuancheng and Ju, Zeqian and Tan, Xu and He, Lei and Wu, Zhizheng and Bian, Jiang and others},
  journal={Advances in Neural Information Processing Systems},
  volume={36},
  pages={71340--71357},
  year={2023}
}

@article{manor2024zero,
  title={Zero-shot unsupervised and text-based audio editing using DDPM inversion},
  author={Manor, Hila and Michaeli, Tomer},
  journal={arXiv preprint arXiv:2402.10009},
  year={2024}
}

@inproceedings{jia2025audioeditor,
  title={Audioeditor: A training-free diffusion-based audio editing framework},
  author={Jia, Yuhang and Chen, Yang and Zhao, Jinghua and Zhao, Shiwan and Zeng, Wenjia and Chen, Yong and Qin, Yong},
  booktitle={ICASSP 2025-2025 IEEE International Conference on Acoustics, Speech and Signal Processing (ICASSP)},
  pages={1--5},
  year={2025},
  organization={IEEE}
}

@inproceedings{cheng2025omnisep,
  title={Omnisep: Unified omni-modality sound separation with query-mixup},
  author={Cheng, Xize and Zheng, Siqi and Fang, Minghui and Zhang, Ziang and Huang, Rongjie and Ji, Shengpeng and Zuo, Jialong and Jin, Tao and Zhao, Zhou and others},
  booktitle={International Conference on Learning Representations},
  volume={2025},
  pages={55196--55213},
  year={2025}
}

@article{liang2025audiomorphix,
  title={AudioMorphix: Training-free audio editing with diffusion probabilistic models},
  author={Liang, Jinhua and Chen, Yuanzhe and Yuan, Yi and Jia, Dongya and Zhuang, Xiaobin and Chen, Zhuo and Wang, Yuping and Wang, Yuxuan},
  journal={arXiv preprint arXiv:2505.16076},
  year={2025}
}

@inproceedings{takahashi2026mmaudiosep,
  title={MMAudioSep: Taming video-to-audio generative model towards video/text-queried sound separation},
  author={Takahashi, Akira and Takahashi, Shusuke and Mitsufuji, Yuki},
  booktitle={ICASSP 2026-2026 IEEE International Conference on Acoustics, Speech and Signal Processing (ICASSP)},
  pages={15667--15671},
  year={2026},
  organization={IEEE}
}

@article{ungersbock2026sao,
  title={Sao-instruct: Free-form audio editing using natural language instructions},
  author={Ungersb{\"o}ck, Michael and Gr{\"o}tschla, Florian and Lanzend{\"o}rfer, Luca and Yi, June Young and Choi, Changho and Wattenhofer, Roger},
  journal={Advances in Neural Information Processing Systems},
  volume={38},
  pages={83411--83437},
  year={2026}
}

@article{shi2025sam,
  title={Sam audio: Segment anything in audio},
  author={Shi, Bowen and Tjandra, Andros and Hoffman, John and Wang, Helin and Wu, Yi-Chiao and Gao, Luya and Richter, Julius and Le, Matt and Vyas, Apoorv and Chen, Sanyuan and others},
  journal={arXiv preprint arXiv:2512.18099},
  year={2025}
}

@article{tao2025mmedit,
  title={MMEDIT: A Unified Framework for Multi-Type Audio Editing via Audio Language Model},
  author={Tao, Ye and Wu, Wen and Zhang, Chao and Wu, Mengyue and Wang, Shuai and Xu, Xuenan},
  journal={arXiv preprint arXiv:2512.20339},
  year={2025}
}

@article{tian2026audio,
  title={Audio-Omni: Extending Multi-modal Understanding to Versatile Audio Generation and Editing},
  author={Tian, Zeyue and Yang, Binxin and Liu, Zhaoyang and Zhang, Jiexuan and Yuan, Ruibin and Yin, Hubery and Chen, Qifeng and Li, Chen and Lyu, Jing and Xue, Wei and others},
  journal={arXiv preprint arXiv:2604.10708},
  year={2026}
}

@article{li2026unison,
  title={UNISON: A Unified Sound Generation and Editing Framework via Deep LLM Fusion},
  author={Li, Zhaoqing and Xu, Haoning and Su, Jingran and Liu, Yaofang and Rao, Zhefan and Wang, Huimeng and Deng, Jiajun and Wang, Tianzi and Jin, Zengrui and Liu, Rui and others},
  journal={arXiv preprint arXiv:2605.31530},
  year={2026}
}

@article{ge2026directaudioedit,
  title={DirectAudioEdit: Inversion-Free Text-Guided Audio Editing via Diffusion Prediction Contrast},
  author={Ge, Zhengkun and Liu, Xiaoqian and Zhang, Haoran and Ge, Yuan and Zhang, Junxiang and Yu, Zhengtao and Zhu, Jingbo and Xiao, Tong},
  journal={arXiv preprint arXiv:2606.07356},
  year={2026}
}

@article{dong2026unified,
  title={Unified Audio Generation and Editing via Joint Condition Modeling and Progressive Training},
  author={Dong, Haocheng and Lu, Yuheng and Gong, Cheng and Liu, Shansong and Zhang, Xiao-Lei and Li, Xuelong},
  journal={arXiv preprint arXiv:2606.16435},
  year={2026}
}

@misc{xu2025qwen3omnitechnicalreport,
      title={Qwen3-Omni Technical Report}, 
      author={Jin Xu and Zhifang Guo and Hangrui Hu and Yunfei Chu and Xiong Wang and Jinzheng He and Yuxuan Wang and Xian Shi and Ting He and Xinfa Zhu and Yuanjun Lv and Yongqi Wang and Dake Guo and He Wang and Linhan Ma and Pei Zhang and Xinyu Zhang and Hongkun Hao and Zishan Guo and Baosong Yang and Bin Zhang and Ziyang Ma and Xipin Wei and Shuai Bai and Keqin Chen and Xuejing Liu and Peng Wang and Mingkun Yang and Dayiheng Liu and Xingzhang Ren and Bo Zheng and Rui Men and Fan Zhou and Bowen Yu and Jianxin Yang and Le Yu and Jingren Zhou and Junyang Lin},
      year={2025},
      eprint={2509.17765},
      archivePrefix={arXiv},
      primaryClass={cs.CL}
}

@inproceedings{laionclap2023,
  title={Large-scale contrastive language-audio pretraining with feature fusion and keyword-to-caption augmentation},
  author={Wu, Yusong and Chen, Ke and Zhang, Tianyu and Hui, Yuchen and Berg-Kirkpatrick, Taylor and Dubnov, Shlomo},
  booktitle={ICASSP 2023-2023 IEEE International Conference on Acoustics, Speech and Signal Processing (ICASSP)},
  pages={1--5},
  year={2023},
  organization={IEEE}
}

@misc{wang2026samaudiojudgeunified,
      title={SAM Audio Judge: A Unified Multimodal Framework for Perceptual Evaluation of Audio Separation}, 
      author={Helin Wang and Bowen Shi and Andros Tjandra and John Hoffman and Yi-Chiao Wu and Apoorv Vyas and Najim Dehak and Ann Lee and Wei-Ning Hsu},
      year={2026},
      eprint={2601.19702},
      archivePrefix={arXiv},
      primaryClass={eess.AS}
}

@inproceedings{chen2020vggsound,
	title={Vggsound: A large-scale audio-visual dataset},
	author={Chen, Honglie and Xie, Weidi and Vedaldi, Andrea and Zisserman, Andrew},
	booktitle={ICASSP 2020-2020 IEEE International Conference on Acoustics, Speech and Signal Processing (ICASSP)},
	pages={721--725},
	year={2020},
	organization={IEEE}
}

@inproceedings{gemmeke2017audio,
  title={AudioSet: An ontology and human-labeled dataset for audio events},
  author={Gemmeke, Jort F and Ellis, Daniel PW and Freedman, Dylan and Jansen, Aren and Lawrence, Wade and Moore, R Channing and Plakal, Manoj and Ritter, Marvin},
  booktitle={2017 IEEE international conference on acoustics, speech and signal processing (ICASSP)},
  pages={776--780},
  year={2017},
  organization={IEEE}
}

@inproceedings{niu2026acavcaps,
  title={ACAVCaps: Enabling large-scale training for fine-grained and diverse audio understanding},
  author={Niu, Yadong and Wang, Tianzi and Dinkel, Heinrich and Sun, Xingwei and Zhou, Jiahao and Li, Gang and Liu, Jizhong and Zhang, Junbo and Luan, Jian},
  booktitle={ICASSP 2026-2026 IEEE International Conference on Acoustics, Speech and Signal Processing (ICASSP)},
  pages={15347--15351},
  year={2026},
  organization={IEEE}
}

@article{mei2024wavcaps,
  title={WavCaps: A ChatGPT-Assisted Weakly-Labelled Audio Captioning Dataset for Audio-Language Multimodal Research},
  author={Mei, Xinhao and Meng, Chutong and Liu, Haohe and Kong, Qiuqiang and Ko, Tom and Zhao, Chengqi and Plumbley, Mark D and Zou, Yuexian and Wang, Wenwu},
  journal={IEEE/ACM Transactions on Audio, Speech, and Language Processing},
  volume={32},
  pages={3339--3354},
  year={2024},
  publisher={IEEE}
}

@article{tjandra2025meta,
  title={Meta audiobox aesthetics: Unified automatic quality assessment for speech, music, and sound},
  author={Tjandra, Andros and Wu, Yi-Chiao and Guo, Baishan and Hoffman, John and Ellis, Brian and Vyas, Apoorv and Shi, Bowen and Chen, Sanyuan and Le, Matt and Zacharov, Nick and others},
  journal={arXiv preprint arXiv:2502.05139},
  year={2025}
}

@InProceedings{liu2024tackling,
    author    = {Liu, Xiulong and Dong, Zhikang and Zhang, Peng},
    title     = {Tackling Data Bias in MUSIC-AVQA: Crafting a Balanced Dataset for Unbiased Question-Answering},
    booktitle = {Proceedings of the IEEE/CVF Winter Conference on Applications of Computer Vision (WACV)},
    month     = {January},
    year      = {2024},
    pages     = {4478-4487}
}

@article{ephrat2018looking,
   title={Looking to listen at the cocktail party: a speaker-independent audio-visual model for speech separation},
   volume={37},
   ISSN={1557-7368},
   url={http://dx.doi.org/10.1145/3197517.3201357},
   DOI={10.1145/3197517.3201357},
   number={4},
   journal={ACM Transactions on Graphics},
   publisher={Association for Computing Machinery (ACM)},
   author={Ephrat, Ariel and Mosseri, Inbar and Lang, Oran and Dekel, Tali and Wilson, Kevin and Hassidim, Avinatan and Freeman, William T. and Rubinstein, Michael},
   year={2018},
   month=July, pages={1–11} }

@inproceedings{bain2020condensed,
  booktitle = {15th Asian Conference on Computer Vision, 2020},
  pages = {460-479},
  publisher = {Springer},
  title = {Condensed movies: story based retrieval with contextual embeddings},
  author = {Bain, M and Nagrani, A and Brown, A and Zisserman, A},
  year = {2021},
  organizer = {15th Asian Conference on Computer Vision, 2020},
  series = {Lecture Notes in Computer Science}
}

@inproceedings{zhou2022audio,
  title={Audio--visual segmentation},
  author={Zhou, Jinxing and Wang, Jianyuan and Zhang, Jiayi and Sun, Weixuan and Zhang, Jing and Birchfield, Stan and Guo, Dan and Kong, Lingpeng and Wang, Meng and Zhong, Yiran},
  booktitle={European Conference on Computer Vision},
  pages={386--403},
  year={2022},
  organization={Springer}
}

@article{seedance2026seedance,
  title={Seedance 2.0: Advancing video generation for world complexity},
  author={Seedance, Team and Chen, De and Chen, Liyang and Chen, Xin and Chen, Ying and Chen, Zhuo and Chen, Zhuowei and Cheng, Feng and Cheng, Tianheng and Cheng, Yufeng and others},
  journal={arXiv preprint arXiv:2604.14148},
  year={2026}
}

@misc{gong2021ast,
      title={AST: Audio Spectrogram Transformer}, 
      author={Yuan Gong and Yu-An Chung and James Glass},
      year={2021},
      eprint={2104.01778},
      archivePrefix={arXiv},
      primaryClass={cs.SD},
      url={https://arxiv.org/abs/2104.01778}, 
}

@article{kong2020panns,
  title={Panns: Large-scale pretrained audio neural networks for audio pattern recognition},
  author={Kong, Qiuqiang and Cao, Yin and Iqbal, Turab and Wang, Yuxuan and Wang, Wenwu and Plumbley, Mark D},
  journal={IEEE/ACM Transactions on Audio, Speech, and Language Processing},
  volume={28},
  pages={2880--2894},
  year={2020},
  publisher={IEEE}
}

@inproceedings{girdhar2023imagebind,
  title={Imagebind: One embedding space to bind them all},
  author={Girdhar, Rohit and El-Nouby, Alaaeldin and Liu, Zhuang and Singh, Mannat and Alwala, Kalyan Vasudev and Joulin, Armand and Misra, Ishan},
  booktitle={Proceedings of the IEEE/CVF conference on computer vision and pattern recognition},
  pages={15180--15190},
  year={2023}
}

@article{comanici2025gemini,
  title={Gemini 2.5: Pushing the frontier with advanced reasoning, multimodality, long context, and next generation agentic capabilities},
  author={Comanici, Gheorghe and Bieber, Eric and Schaekermann, Mike and Pasupat, Ice and Sachdeva, Noveen and Dhillon, Inderjit and Blistein, Marcel and Ram, Ori and Zhang, Dan and Rosen, Evan and others},
  journal={arXiv preprint arXiv:2507.06261},
  year={2025}
}

@article{ma2026mmae,
  title={MMAE: A Massive Multitask Audio Editing Benchmark},
  author={Ma, Ziyang and Yan, Ruiqi and Xu, Ruiyang and Fang, Jie and Niu, Zhikang and Chao, Yi-Wen and Tu, Wenming and Wang, Tianrui and Chen, Qi and Chen, Wenxi and others},
  journal={arXiv preprint arXiv:2606.07229},
  year={2026}
}

\clearpage
\section{Appendix}
\label{sec:supp}

\subsection{Dataset Annotation and Quality Filtering}
\label{sec:A}

\begin{figure}[h]
    \centering
    \includegraphics[width=1.0\linewidth]{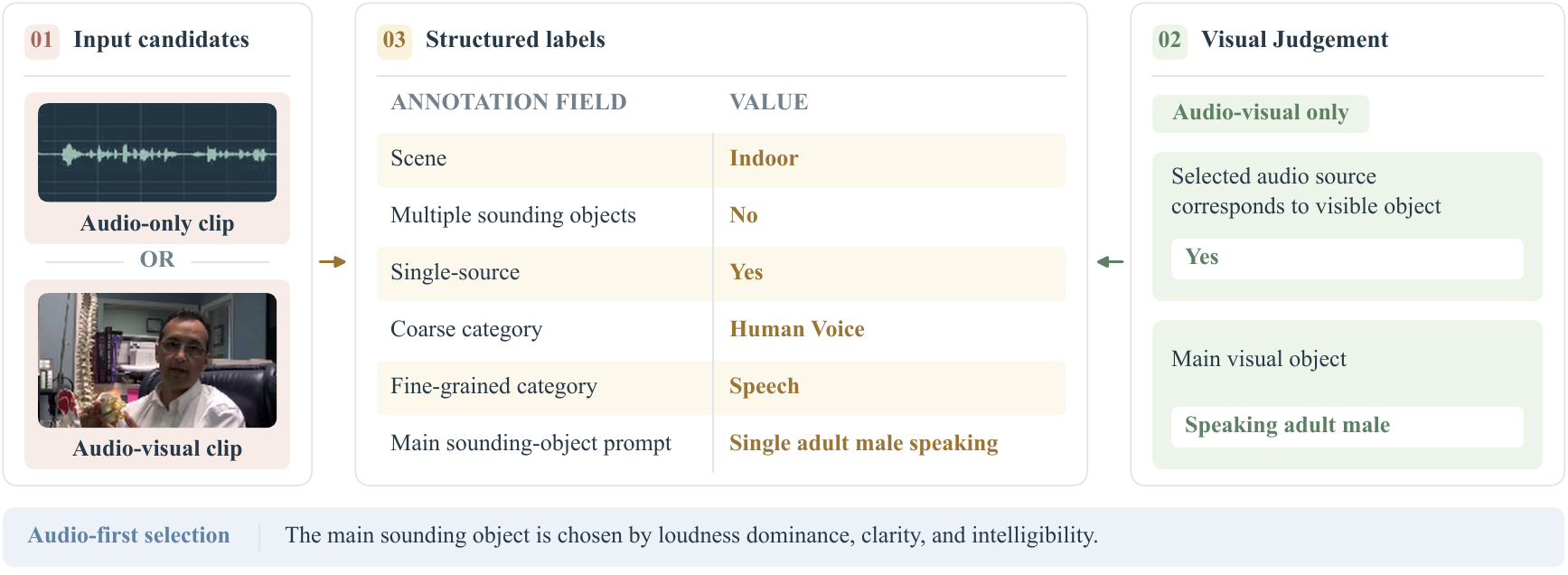}
    \caption{Qwen3-Omni Annotation Example.}
    \label{fig:qwen3-annotation}
\end{figure}

\begin{figure}[t]
    \centering
    \includegraphics[width=\linewidth]{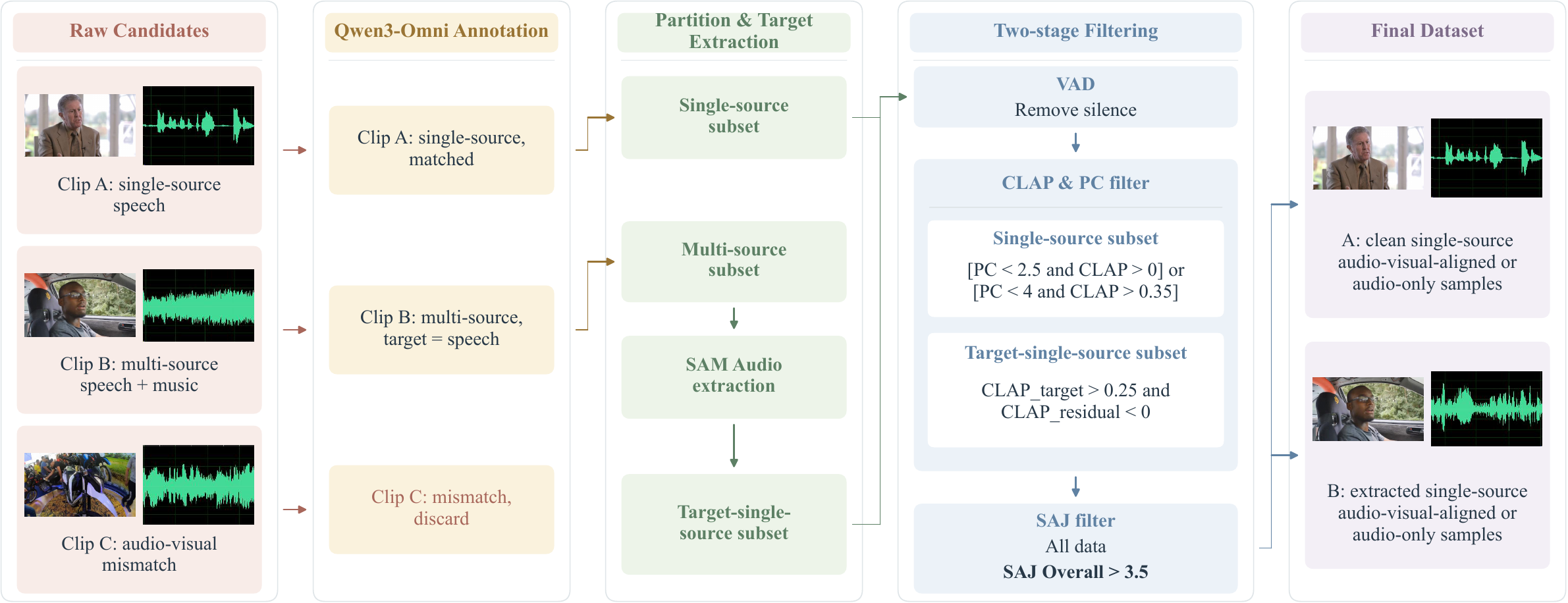}
    \caption{Data Construction Example.}
    \label{fig:data-construction}
\end{figure}

We use Qwen3-Omni to obtain structured labels for raw audio-only and audio-visual candidates. The protocol is audio-first: the main sounding object is selected according to the audio evidence rather than the visible subject. Before producing the final label, the model is instructed to internally compare candidate sources based on loudness dominance, clarity, and intelligibility, and to choose a single dominant source.

For audio-only samples, Qwen3-Omni outputs six fields: scene type, whether multiple similar sounding objects exist, whether the clip is effectively single-source for the selected main object, the coarse sound category, the fine-grained category, and a concise main sounding-object prompt. The scene label is selected from \{Indoor, Urban, Nature\}. The multiple-source and single-source judgments are binary Yes/No labels. The coarse category is selected from \{Human Voice, Music, SFX\}, while the fine-grained category is selected from \{Speech, Non-verbal, Crowd, Instrument, Score, Animal, Ambience, Mechanical, Foley\}. The prompt is written as a short NP- or VP-style phrase, such as ``dog barking'', ``single adult male speaking'', or ``piano playing''. 
For audio-visual samples, Qwen3-Omni outputs the same audio-centric fields and additionally predicts whether the selected audio source corresponds to the main visible object and what the main visual object is. The visual correspondence label is set to Yes only when the dominant sounding object is clearly visible, visually salient, and plausibly responsible for the dominant sound. Otherwise, the sample is regarded as audio-visual mismatched and rejected from the aligned data pool. This rule prevents visually salient but acoustically irrelevant objects from being used as positive audio-visual pairs. 
The output space is therefore fixed and auditable: binary fields use Yes/No labels, scene and sound types are selected from predefined category sets, and the sounding-object prompt is constrained to a concise and standalone NP- or VP-style phrase. 

After data annotation, the pipeline first filters out audio-video samples where the sounding object does not match the visual content. All audio-video and audio-only samples are then split into single-source and multi-source subsets according to the binary single-source label. Multi-source samples containing multiple similar sounding objects are discarded. For remaining multi-source samples, SAM Audio extracts target audio guided by the prompt of the main sounding object; these extracted audio samples are fed into the filtering module together with raw single-source samples. The first filtering stage removes silent segments via VAD detection. Next, CLAP (audio-text semantic alignment via embedding cosine similarity) and PC (Production Complexity, quantifying acoustic scene complexity) scores are jointly adopted to evaluate candidate samples in terms of semantic alignment and structural complexity. For native single-source samples, we retain samples satisfying either (PC < 2.5 and CLAP > 0) or (PC < 4 and CLAP > 0.35). For samples derived from multi-source data after target audio extraction by SAM Audio, we keep cases where the CLAP score between SAM-extracted target audio and the target label exceeds 0.25, while the CLAP score between residual audio and the target label is below 0. The second filtering stage leverages SAJ scores to screen samples based on audio quality. Specifically, only samples with an SAJ score higher than 3.5 are preserved. Through the above pipeline, around 50\% of candidate samples are eliminated, yielding approximately one million high-quality single-target audio-video aligned samples. 
Figure~\ref{fig:qwen3-annotation} and Figure~\ref{fig:data-construction} illustrate the annotation example generated by Qwen3-Omni and the complete dataset construction pipeline.

\subsection{Instruction Templates}
\label{sec:B}

Our training employs generalized instructions to cover both task-level and language-level diversity. At the task level, templates are organized into extraction, deletion, and combined preserve-and-remove modes, so that the model learns positive preservation, negative suppression, and their joint use in a single command. At the language level, each mode is expressed by multiple surface forms to reduce prompt overfitting and improve robustness to natural user instructions.

For extraction, the target source is denoted as $x$, and we use the following templates: ``extract $x$'', ``isolate $x$'', ``separate $x$'', ``keep $x$'', ``keep only $x$'', ``only keep $x$'', ``I want to hear $x$'', ``I want only $x$'', ``give me $x$'', ``give me only $x$'', ``leave only $x$'', ``preserve $x$'', ``retain $x$'', ``the sound of $x$'', ``focus on $x$'', ``highlight $x$'', ``amplify $x$'', ``let me hear $x$'', ``please isolate $x$'', and ``please keep $x$''.

For deletion, the removed source is denoted as $y$, and we use the following templates: ``delete $y$ and keep all other sounds'', ``remove $y$ and keep all other sounds'', ``eliminate $y$ and preserve the remaining audio'', ``exclude $y$ while preserving everything else'', ``get rid of $y$ but keep the other sounds'', ``suppress $y$ and keep the background intact'', ``mute $y$ without changing the other sounds'', ``silence $y$ and preserve all other audio'', ``drop $y$ while keeping the remaining sounds'', ``discard $y$ without removing anything else'', ``remove the sound of $y$ and keep everything else'', ``I do not want $y$, but keep the rest'', ``do not include $y$, preserve all other sounds'', ``without $y$, keep the remaining audio unchanged'', ``no $y$, keep all other sounds'', ``everything except $y$'', ``anything but $y$'', ``all sounds except $y$'', ``please remove only $y$'', and ``kindly remove $y$ and keep the rest''.

For combined preserve-and-remove instructions, we first sample an extraction phrase $\mathrm{ext}(x)$ from the extraction templates and a removal phrase $\mathrm{rem}(y)$ from a deletion-oriented phrase pool, including ``delete $y$'', ``remove $y$'', ``eliminate $y$'', ``exclude $y$'', ``get rid of $y$'', ``suppress $y$'', ``mute $y$'', ``silence $y$'', ``drop $y$'', ``discard $y$'', ``remove the sound of $y$'', ``I do not want $y$'', ``do not include $y$'', ``without $y$'', ``no $y$'', ``please remove $y$'', and ``kindly remove $y$''. The two phrases are then composed with connective templates: ``$\mathrm{ext}(x)$ and $\mathrm{rem}(y)$'', ``$\mathrm{ext}(x)$, and $\mathrm{rem}(y)$'', ``$\mathrm{ext}(x)$; $\mathrm{rem}(y)$'', ``$\mathrm{ext}(x)$ but $\mathrm{rem}(y)$'', ``$\mathrm{ext}(x)$ while $\mathrm{rem}(y)$'', ``$\mathrm{rem}(y)$ and $\mathrm{ext}(x)$'', ``$\mathrm{rem}(y)$, then $\mathrm{ext}(x)$'', ``$\mathrm{rem}(y)$ so I can hear $x$'', and ``$\mathrm{ext}(x)$ and at the same time $\mathrm{rem}(y)$''. During training, textual surface perturbations such as capitalization, punctuation, and polite suffixes are randomly applied.

\subsection{Multi-Task Training and Mixture Construction}
\label{sec:C}

\begin{figure}[t]
    \centering
    \includegraphics[width=1.0\linewidth]{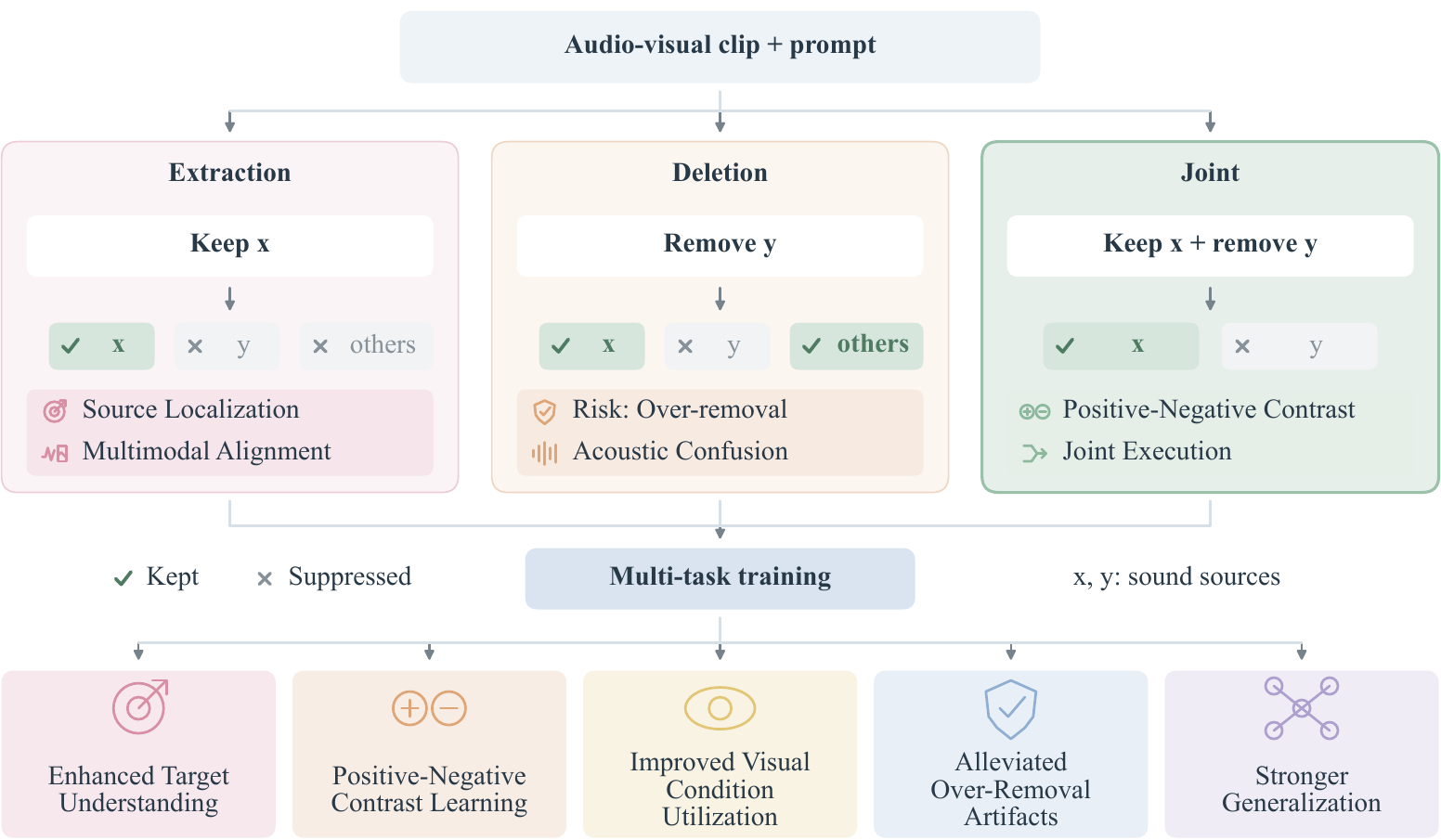}
    \caption{Multi-Task Training Paradigm.}
    \label{fig:multi-task-training}
\end{figure}

Multi-task training complements the deletion objective with extraction and joint editing samples. Extraction improves target-source localization and multimodal condition alignment, while joint editing exposes the model to positive preservation and negative suppression in the same instruction. This design reduces over-dependence on deletion-only prompts and helps prevent over-removal of acoustically similar non-target sources. Specifically, multi-task training yields several distinct benefits, as summarized below.
\begin{itemize}
\item \textbf{Enhanced target understanding.} The extraction task forces the model to learn the explicit correspondence between visual/textual conditions and sounding sources. With such awareness, the model can accurately determine which acoustic components should be removed during deletion-oriented editing.
\item \textbf{Positive-negative contrast learning.} Joint optimization of two opposite task polarities enables the model to clearly distinguish between preservable and suppressible acoustic targets.
\item \textbf{Improved visual condition utilization.} Since visual cues primarily describe objects to be retained, the extraction task strengthens the binding between visual objects and their corresponding audio components. This avoids the model's over-reliance on textual conditions in pure deletion scenarios.
\item \textbf{Alleviated over-removal artifacts.} Models trained solely on deletion tasks tend to indiscriminately suppress background sounds and acoustically similar interfering sources. Incorporating extraction supervision guides the model to produce structurally meaningful and source-aware outputs.
\item \textbf{Stronger generalization capability.} Introducing extraction and composite preserve-and-remove tasks enriches the semantic coverage of training instructions, leading to better generalization toward diverse real-world user prompts.
\end{itemize} 
Figure~\ref{fig:multi-task-training} depicts the proposed multi-task training paradigm with extraction, deletion and joint preserve-and-remove tasks, and summarizes the corresponding advantages of joint training.

\begin{figure}[t]
    \centering
    \includegraphics[width=1.0\linewidth]{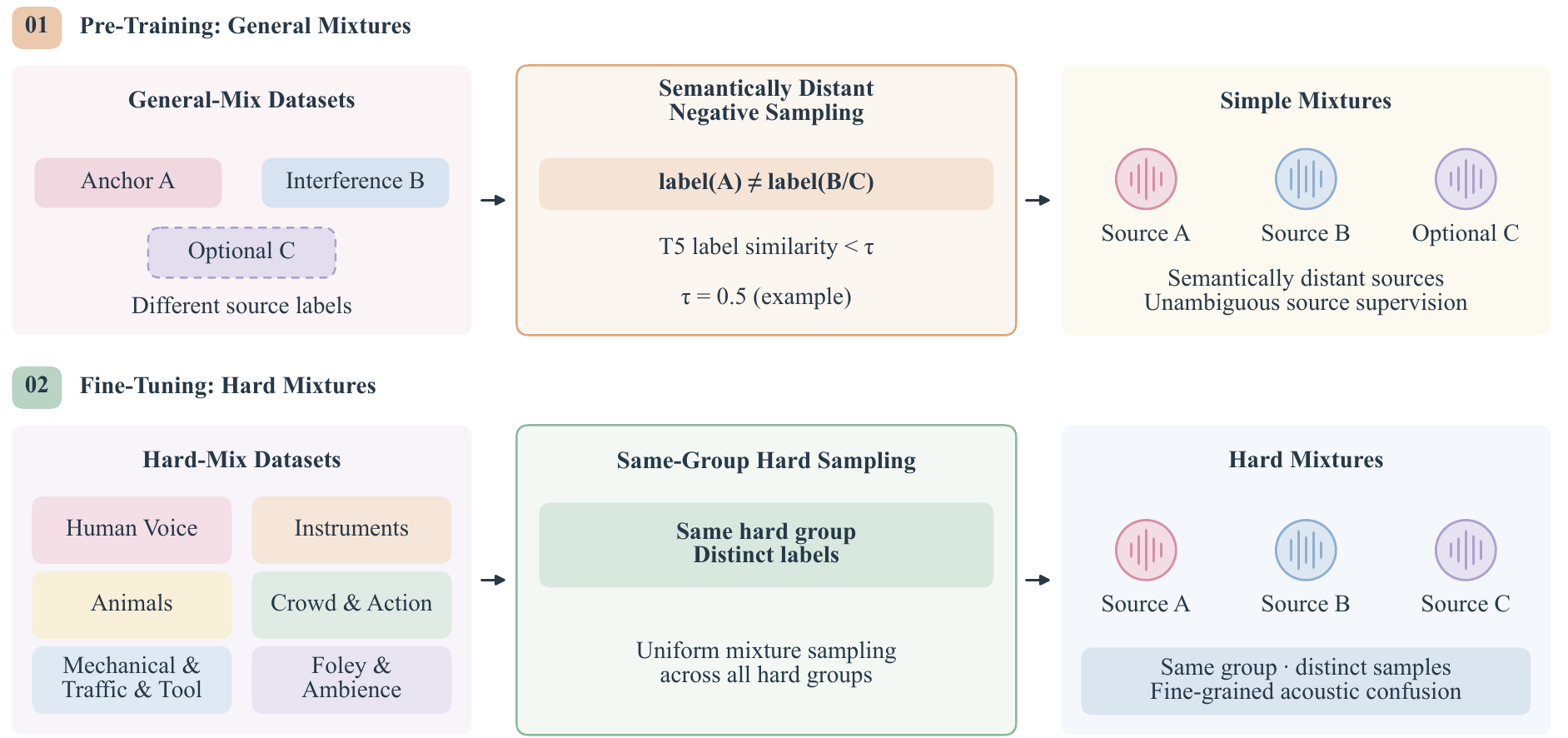}
    \caption{Two-Stage Curriculum Sampling Paradigm.}
    \label{fig:two-stage-training}
\end{figure}

Training samples for the two-stage curriculum are divided into easy and hard samples according to different training phases. In the pre-training stage, simple mixtures are sampled from general-mix datasets using semantically-far negative sampling: For an anchor source $A$, interference sources $B$ and optional interference sources $C$ must have labels different from $A$, and their T5-based label similarity to the selected sources must be lower than a preset threshold (e.g., 0.5). This generates semantically distant source pairs and provides unambiguous supervision for source discrimination. 
In the fine-tuning stage, difficult mixtures are sampled from hard-mix datasets. These hard-mix datasets contain groups of samples with fine-grained acoustically confusing sounds. The defined groups include human voice groups (\textit{\allowbreak voice\_age\_emotion}, \textit{\allowbreak voice\_gender\_speech}, \textit{\allowbreak voice\_gender\_singing}) covering male/female speech, male/female singing, children crying, laughter, screams, crowd sounds and more; instrument groups (\textit{\allowbreak instruments\_strings}, \textit{\allowbreak instruments\_winds}, \textit{\allowbreak instruments\_percussion}) including strings, winds, and percussion; animal sound groups (\textit{\allowbreak animals\_mammals\_pets}, \textit{\allowbreak animals\_birds}, \textit{\allowbreak animals\_insects\_amphibians}) consisting of mammals such as cats, dogs, cattle, sheep, horses, donkeys and wolves, birds like ducks, geese, crows and parrots, as well as insects and frogs; crowd and action groups; mechanical, traffic and tool groups; foley material groups; and natural ambience groups. During the fine-tuning stage, each training mixture is constructed by mixing distinct samples drawn from the same hard group, and mixtures are uniformly sampled across all hard groups. Figure~\ref{fig:two-stage-training} illustrates curriculum sampling from semantically distant mixtures in pre-training to acoustically confusing, same-group mixtures in fine-tuning.

\subsection{AV-Remove-Bench Design}
\label{sec:D}

AV-Remove-Bench is designed to evaluate whether an audio editing model can make the soundtrack consistent with a visually edited video. Each sample is an audio-visual clip of about six seconds and specifies one sounding object to be removed from both the visual and audio streams. The benchmark contains two scenario types: multi-sounding-object scenes, where a single clip contains two or more clearly distinguishable sounding objects and corresponding audio sources, and foreground-target-plus-stable-background scenes, where a foreground target sound is mixed with a stable background ambience.

The benchmark draws from three data sources. Public-dataset samples cover music, speech, and sound effects: Music-AVQA and Music-Duets are used for music cases, AVSpeech for speech cases, and Condensed Movies, AVSBench, and VGGSound-test for sound-effect cases. Generated samples are produced by Seedance 2.0 from manually designed prompts. These prompts are organized to cover speech, music, and sound effects, with sound effects further spanning tool actions, natural ambience, electronic and mechanical sounds, vehicles, and animals. The prompt set also explicitly balances the two scenario types above, so that generated clips include both multiple foreground sounding objects and foreground objects accompanied by stable background sound. Real recordings are collected from daily real-world scenes and mainly focus on sound-effect-centric object removal, such as vehicles and environmental sounds.

For public datasets, we manually select 42 high-quality samples from the aforementioned sources. For synthetic samples, we adopt the top 27 generated instances. For real recordings, we carefully collect samples covering eight distinct real-life scenarios with clear documentation. Furthermore, an example prompt for Seedance 2.0 video generation is provided below:
\begin{lstlisting}
Two people working on a backyard fence --- one hammering nails with a hammer, the other sawing a board with a hand saw --- both working simultaneously. Landscape orientation (16:9 horizontal frame). Fixed camera, side view capturing both workers beside the fence. AUDIO REQUIREMENTS: Two clearly distinguishable tool sounds emitted by two different visible people working concurrently: (1) sharp metallic clang produced by the hammer striking nails into wood; (2) rhythmic back-and-forth rasping sound as the saw cuts through the board. The two sounds overlap continuously, and each sound is synchronized with the visible motions of the corresponding worker. Fixed camera, clip duration of 6 seconds. No voiceover, narration, or off-screen audio. High signal-to-noise ratio, clear and crisp audio, high-definition video.
\end{lstlisting}

\begin{figure}[t]
    \centering
    \includegraphics[width=0.75\linewidth]{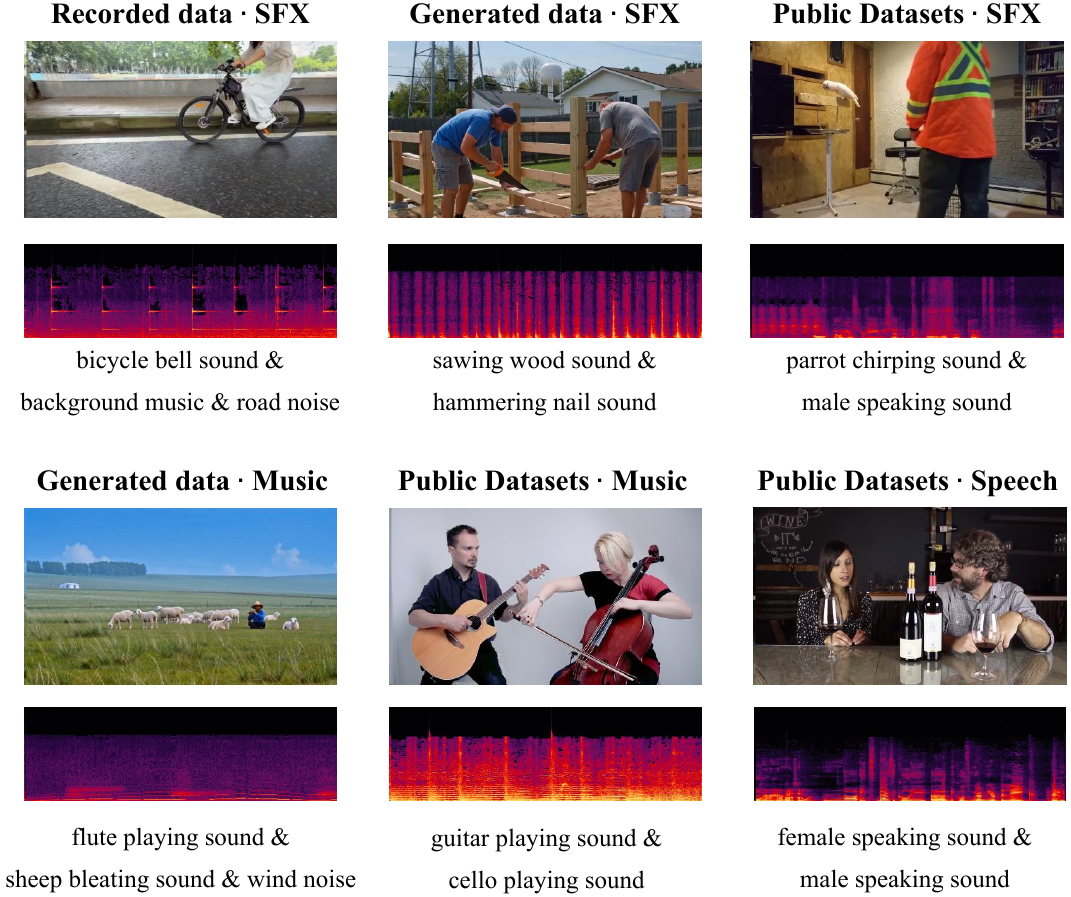}
    \caption{Data examples from AV-Remove-Bench.}
    \label{fig:data_examples}
\end{figure}

Finally, Figure~\ref{fig:data_examples} shows several data examples from AV-Remove-Bench.

\subsection{Gemini-Based Scoring Criteria}
\label{sec:E}

We use Gemini as a multimodal judge to provide subjective-style scores for the edited results. Each evaluation item contains the original video, the edited video, the removal instruction, and, when available, the edit mask. The judge outputs four 1--5 scores: visual target removal (Instruction Compliance$_v$), video fidelity (Fidelity$_v$), audio target removal (Instruction Compliance$_a$), and preserved-audio fidelity (Fidelity$_a$).

For video evaluation, Gemini is instructed to watch both the original and edited videos and to focus only on visual content. Instruction Compliance$_v$ measures whether the target object or action specified by the instruction has disappeared from the edited video. The mask, when provided, is used to localize the edited region. The score is based on semantic removal: changing the target's appearance, color, texture, or blur level does not count as removal if the target identity remains recognizable. Fidelity$_v$ measures both the realism of the generated content inside the mask and the preservation of non-target content outside the mask; severe inpainting artifacts, boundary seams, blur, or damage to preserved objects lower this score.

For audio evaluation, Gemini compares the original and edited audio tracks. Instruction Compliance$_a$ only measures whether the target sound is removed, with the target status categorized as absent, faint, or clear. Fidelity$_a$ measures whether non-target sounds remain audible, natural, and close to the original, while also considering overall audio quality. Near-silent or severely distorted edited audio receives a low fidelity score even if the target sound is absent. The two audio scores are intentionally independent, so a result can achieve high target-removal compliance but low fidelity if it removes all sounds or introduces strong artifacts.

\begin{table}[h]
\centering
\begin{tabular}{lc}
\toprule
metric & Kendall $\uparrow$ \\
\midrule
Instruction Compliance$_v$ & 0.7494 \\
Fidelity$_v$ & 0.7113 \\
Instruction Compliance$_a$ & 0.7063 \\
Fidelity$_a$ & 0.7092 \\
\midrule
Overall & 0.7191 \\
\bottomrule
\end{tabular}
\caption{Consistency Analysis Between MLLM Scores and Human Subjective Ratings}
\label{tab:kendall_consistency}
\end{table}

To verify the accuracy of scores generated by the MLLM, we conduct a consistency analysis between model predictions and human subjective ratings on AV-Remove-Bench, quantified via the Kendall rank correlation coefficient. The results are presented in Table~\ref{tab:kendall_consistency}. The Kendall rank correlation coefficient measures the agreement of sample ordering between two sets of scores and ranges from $-1$ to $1$. Higher values indicate stronger consistency in ranking; zero signifies no correlation, while negative values imply opposite ordering. Our results yield Kendall coefficients of 0.7494, 0.7113, 0.7063 and 0.7092 for each metric, with an overall average of 0.7191. This demonstrates strong agreement between model and human evaluations, validating the reliability of the MLLM.

\subsection{Comparison with Commercial Models}
\label{sec:F}

Seedance 2.0, Seedance 2.5, and MiniMax H3 are commercial closed-source and open-source generative models that support joint audio-video editing. These models enable collaborative modification, regional inpainting and content completion for input videos and their audio tracks conditioned on text or multimodal prompts. We compare our method with these commercial models in terms of target sound removal on AV-Remove-Bench, and the corresponding results are summarized in Table~\ref{tab:seedance_objective}.

\renewcommand{\arraystretch}{1}
\begin{table*}[h]
\centering
\resizebox{\linewidth}{!}{
\begin{tabular}{lcccccccc}
\toprule
Method & \multicolumn{6}{c}{Audio metrics} & \multicolumn{2}{c}{Audio-visual metrics} \\
\cmidrule(lr){2-7} \cmidrule(lr){8-9}
& IS $\uparrow$ & SAJ Overall $\uparrow$ & TSSR $\uparrow$ & PSF $\uparrow$ & Instr. Comp.$_a\uparrow$ & Fidelity$_a\uparrow$ & IB-AV $\uparrow$ & DeSync $\downarrow$ \\
\midrule
Seedance 2.0 & 3.36 & 2.50 & -5.63 & 9.70 & 3.17 & 3.83 & 23.71 & 0.58 \\
Seedance 2.5 & 2.85 & 2.93 & 3.63 & 10.59 & 4.36 & 4.11 & \textbf{30.66} & \textbf{0.49} \\
MiniMax H3 & 3.30 & 2.25 & 0.76 & 9.97 & 3.41 & \textbf{4.30} & 23.52 & 0.64 \\
\textbf{TV-AudioRemover} & \textbf{3.55} & \textbf{3.46} & \textbf{6.35} & \textbf{10.79} & \textbf{4.79} & 3.96 & 27.32 & 0.60 \\
\bottomrule
\end{tabular}}
\caption{Objective Evaluation: TV-AudioRemover versus commercial models on the AV-Remove-Bench.}
\label{tab:seedance_objective}
\end{table*}
\renewcommand{\arraystretch}{1}

The results demonstrate that our model achieves overall superior performance on the target sound removal task compared with these commercial models. Seedance 2.0 and MiniMax H3 exhibit limited instruction-following capability in audio editing. Seedance 2.5 achieves better audio-editing performance than Seedance 2.0, yet it tends to fully regenerate both video frames and audio tracks. This behavior yields strong audio-visual synchronization at the cost of altering non-target content.

An example prompt for audio-visual editing with Seedance 2.0, Seedance 2.5 and MiniMax H3 is provided below:
\begin{lstlisting}
Perform integrated editing for both video frames and audio tracks synchronously:
Visual editing: Remove all subjects that produce meowing sounds (the meowing cat) from the frames. The barking dog, dog barks, all other sound sources and visual content should be fully preserved without any deletion, replacement or modification.
Audio editing: Completely eliminate all cat meows. The dog barks, the barking dog and all other sound sources are retained. The generated result requires natural audiovisual temporal alignment, free of audiovisual desynchronization and unnatural acoustic artifacts.
\end{lstlisting}

\subsection{Objective Metric Computation}
\label{sec:G}

We adopt two groups of objective metrics. The first group is model-based and quantifies acoustic quality, audio-visual consistency, target suppression, and preserved-source fidelity. The second group is MLLM-based and assesses whether the edited outputs follow the removal instruction while preserving non-target content.

For audio quality, IS is computed with a pre-trained PANNs audio classifier and summarizes both the confidence and diversity of predicted audio classes. The classifier outputs probability distributions over audio categories for every edited audio clip, and the average category distribution across the full test set is estimated. We compute the KL divergence between each sample’s predicted distribution and the global distribution. The final IS score is obtained by averaging all KL divergence values and applying an exponential transformation.
SAJ Overall is computed with SAM Audio Judge; it aggregates faithfulness, recall, and precision. Faithfulness measures whether preserved sounds remain undistorted, recall evaluates whether preserved sources stay complete, and precision quantifies whether leakage of unwanted sources is suppressed.

For audio-visual consistency, IB-AV corresponds to the cosine similarity between audio and video embeddings extracted by ImageBind, with higher values indicating stronger cross-modal alignment. DeSync is estimated via Synchformer and measures the predicted temporal offset between the edited audio and video; lower values indicate better synchronization. In Table~\ref{tab:cmp}, for audio-only editing baselines, IB-AV and DeSync are computed after pairing each audio-editing output with SVOR, the best-performing visual removal model. This setup ensures these metrics reflect audio-visual consistency under identical edited-video conditions. The video-side evaluation used to select SVOR is provided in Appendix \cref{sec:H}.

For removal-specific metrics, Rel-TSSR and PSF are computed using an AST audio classifier. Their formulations are presented in the AV-Remove-Bench subsection of the main text.

For MLLM-based evaluation, Gemini takes the original video, edited video, removal instruction, and an optional edit mask as input. Detailed definitions are provided in Appendix \cref{sec:E}.

\subsection{Visual Removal Evaluation}
\label{sec:H}

\begin{table}[h]
\centering
\begin{tabular}{lccc}
\toprule
Method & Instr. Comp.$_v\uparrow$ & Fidelity$_v\uparrow$\\
\midrule
ROSE & 4.86 & 3.30 \\
AVI-Edit & 2.68 & 2.48 \\
EffectErase & 4.74 & 3.33 \\
UnderEraser & 4.91 & 3.49 \\
SVOR & \textbf{5.00} & \textbf{4.03} \\
\bottomrule
\end{tabular}
\caption{Gemini-Based Evaluation of Visual Removal Quality on AV-Remove-Bench.}
\label{tab:supp_video_gemini}
\end{table}

\begin{table*}[h]
\centering
\begin{tabular}{lccccccc}
\toprule
Method & TRC $\uparrow$ & BP $\uparrow$ & BN $\uparrow$ & CR $\uparrow$ & TC $\uparrow$ & AC $\uparrow$ & VSR $\uparrow$ \\
\midrule
ROSE & 0.53 & 0.49 & 0.38 & 0.44 & 0.42 & 46.95 & 42.21\% \\
AVI-Edit & 0.01 & 0.42 & 0.21 & 0.19 & 0.34 & 22.15 & 0.65\% \\
EffectErase & 0.64 & 0.47 & 0.50 & 0.46 & 0.50 & 52.72 & 41.56\% \\
UnderEraser & 0.78 & 0.69 & 0.68 & 0.64 & 0.59 & 70.03 & 68.83\% \\
SVOR & \textbf{0.97} & \textbf{0.91} & \textbf{0.94} & \textbf{0.94} & \textbf{0.98} & \textbf{94.43} & \textbf{91.56\%} \\
\bottomrule
\end{tabular}
\caption{Subjective video-quality evaluation on AV-Remove-Bench.}
\label{tab:supp_video_subjective}
\end{table*}

We select SVOR as the visual removal model for our sound removal model according to the video-side evaluations in Table~\ref{tab:supp_video_gemini} and Table~\ref{tab:supp_video_subjective}. These evaluations compare the open-source models AVI-Edit, ROSE, EffectErase, UnderEraser, and SVOR on the 77-sample AV-Remove-Bench. Gemini-based scores evaluate whether the edited video follows the removal instruction and preserves non-target visual content, while subjective scores measure target removal completeness (TRC), background preservation (BP), boundary naturalness (BN), generated-content realism (CR), and temporal consistency (TC). AC denotes the video composite score, computed as $(0.3\times$TRC $+0.3\times$BP $+0.15\times$BN $+0.15\times$CR$+0.1\times$TC$)\times 100$, and VSR denotes the video success rate, i.e., the fraction of samples whose four human scores are all non-zero. SVOR obtains the best results in both evaluations, so it is used as the fixed edited-video input when computing IB-AV and DeSync for audio-editing-only methods.

\subsection{Subjective Evaluation Statistics}
\label{sec:I}

For human listening evaluation, 10 raters with professional audio-editing experience independently score anonymized model outputs for each sample, and raters are blinded to model identities during annotation. Each sample is rated along four dimensions on a three-point scale of $\{0, 0.5, 1\}$: target removal completeness (TRC), background preservation (BP), temporal naturalness (TN), and overall quality (OQ). TRC measures whether the target sound has been removed, BP assesses whether non-target sounds are preserved, TN evaluates whether the edited audio sounds temporally smooth and natural, and OQ summarizes the overall perceptual quality of the output. A score of 1 indicates a clear success on the corresponding dimension, 0.5 indicates partial or moderate success, and 0 indicates a clear failure.

\begin{table}[h]
\centering
\begin{tabular}{lccc}
\toprule
\multicolumn{4}{l}{\textbf{GSB scores for TV-AudioRemover versus baselines}}\\
Metric & Good rate & Same rate & Bad rate \\
\midrule
TRC & 0.92 & 0.06 & 0.03 \\
BP & 0.71 & 0.14 & 0.15 \\
TN & 0.86 & 0.10 & 0.04 \\
OQ & 0.67 & 0.21 & 0.13 \\
\bottomrule
\end{tabular}
\captionof{table}{GSB scores for TV-AudioRemover versus baselines.}
\label{tab:user_study_gsb}
\end{table}

For the GSB evaluation, annotators perform pairwise comparisons between TV-AudioRemover and all baseline methods across all samples on each evaluation dimension, and vote for three grades: Better, Same, or Worse. As illustrated in Table~\ref{tab:user_study_gsb}, the GSB pairwise comparison results reveal that TV-AudioRemover obtains considerably higher good rates than bad rates across all evaluation dimensions. Annotators tend to favor the outputs generated by TV-AudioRemover over baseline methods comprehensively, demonstrating the superiority of our approach from human subjective perspectives.

\begin{table}[h]
\centering
\begin{tabular}{lccc}
\toprule
Metric & $p$-value $\downarrow$ & Kendall's $W$ $\uparrow$ & Significant \\
\midrule
TRC & $1.31\times10^{-9}$ & 0.154 & Yes \\
BP & $4.43\times10^{-26}$ & 0.406 & Yes \\
TN & $1.74\times10^{-26}$ & 0.412 & Yes \\
OQ & $6.86\times10^{-26}$ & 0.403 & Yes \\
\bottomrule
\end{tabular}
\caption{Friedman significance test results for subjective metrics on AV-Remove-Bench. All reported metrics exhibit statistically significant differences among the compared models ($p < 0.05$).}
\label{tab:supp_subjective_sig}
\end{table}

To verify whether the observed inter-model gaps in average subjective scores reflect genuine performance differences rather than random sampling noise, we conduct statistical significance tests. The Friedman test is employed to detect overall significant differences across multiple models for each metric, and Kendall’s W quantifies the agreement in rankings among annotators. 
Interpretation of Friedman p-values:
$p < 0.05$ rejects the null hypothesis, confirming significant differences among models;
$p \ge 0.05$ fails to reject the null hypothesis, indicating no statistically significant difference. 
Interpretation of Kendall’s W:
Values near 0 indicate weak ranking consistency across annotators; values close to 1 correspond to high agreement in the rankings provided by different raters. 
The corresponding results are presented in Table~\ref{tab:supp_subjective_sig}. The Friedman test yields $p < 0.05$ for all four subjective evaluation metrics, confirming statistically significant differences among the compared models. These overall statistical results verify that the observed performance gaps in subjective scores are not attributable to random noise.

\subsection{Limitations}
\label{sec:J}

The existing benchmark suffers from limited scale, especially for high-quality sound-effect samples. Larger public benchmarks paired with reference audios after target sound removal will facilitate standardized evaluation. In addition, highly entangled sound sources, such as overlapping speech from visually similar speakers, remain challenging. Such complicated cases demand more powerful modeling of speaker and object identity.

\end{document}